\documentclass[%
aip,
jmp,%
amsmath,amssymb,
reprint,%
]{revtex4-1}

\usepackage{graphicx}
\usepackage{dcolumn}
\usepackage{bm}
\usepackage{graphicx}
\usepackage[displaymath]{lineno}
\usepackage{multirow}
\usepackage{CJKutf8}

\graphicspath{{figs/}}

\usepackage{XiGrpCommon}

\renewcommand{\RevisedText}[1]{#1}

\newcommand{\BibitemShut}[1]{}

\begin{document}

\title{Vortex dynamics in low- and high-extent polymer drag reduction regimes revealed by vortex tracking and conformation analysis}

\begin{CJK*}{UTF8}{gbsn}
\CJKtilde 
\CJKindent 

\author{Lu Zhu\RevisedText{~(朱路)}}
\affiliation{Department of Chemical Engineering, McMaster Universtiy, Hamilton, Ontario L8S 4L7, Canada}
\author{Li Xi\RevisedText{~(奚力)}}
\email[corresponding author, E-mail: ]{xili@mcmaster.ca}

\affiliation{Department of Chemical Engineering, McMaster Universtiy, Hamilton, Ontario L8S 4L7, Canada}
\affiliation{Kavli Institute for Theoretical Physics (KITP), University of California, Santa Barbara, California 93106-4030, U.S.A.}

\date{\today}


\begin{abstract} 
Turbulent flow profiles are known to change between low- (LDR) and high-extent drag reduction (HDR) regimes.
It is however not until recently that the LDR-HDR transition is recognized as a fundamental change between two DR mechanisms.
Although the onset of DR, which initiates the LDR stage, is explainable by a general argument of polymers suppressing vortices, the occurrence of HDR where flow statistics are qualitatively different and DR effects are observed across a much broader range of wall regions remains unexplained.
Recent development of the VATIP (vortex axis tracking by iterative propagation) algorithm allows the detection and extraction of vortex axis-lines with various orientations and curvatures.
This new tool is used in this study to analyze the vortex conformation and dynamics across the LDR-HDR transition.
Polymer effects are shown to concentrate on vortices that are partially or completely attached to the wall.
At LDR, this effect is an across-the-board weakening of vortices which lowers their intensity without shifting their distribution patterns.
At HDR, polymers start to suppress the lift-up of streamwise vortices in the buffer layer and prevent their downstream heads from rising into the log-law layer and forming hairpins and other curved vortices.
This interrupts the turbulent momentum transfer between the buffer and log-law layers, which offers a clear pathway for explaining the distinct mean flow profiles at HDR.
The study depicts the first clear physical picture regarding the changing vortex dynamics between LDR-HDR, which is based on direct evidences from objective statistical analysis of vortex conformation and distribution.
\end{abstract}

\maketitle 
\end{CJK*}

\section{Introduction}\label{Sec_intro}

When a small amount of polymers are added into Newtonian turbulence, their strong interaction with the flow can significantly modify turbulent coherent structures, which results in the drastic reduction of the turbulent friction drag.
Polymer-induced turbulent drag reduction (DR) has been a subject of intense interest
in the literature~\citep{Virk_AIChEJ1975,white2008mechanics,Graham_POF2014} due to its significant practical implications
for the development of flow control techniques for enhanced fluid transportation efficiency.

In polymeric turbulent flows, the Weissenberg number $\mathrm{Wi}\equiv\lambda\dot{\gamma}$ ($\lambda$ and $\dot{\gamma}$ are the polymer relaxation time and the characteristic shear rate of the flow, respectively) measures the level of polymer-induced elasticity.
Polymer effects on turbulent flow statistics are not noticeable until $\mathrm{Wi}$ exceeds a critical magnitude, often denoted by $\mathrm{Wi}_\text{onset}$, which corresponds to the coil-stretch transition of polymer molecules.
After the onset, the level of DR increases with $\mathrm{Wi}$ but
eventually converges to an asymptotic upper bound~\citep{Virk_AIChEJ1975} -- the widely known maximum drag reduction (MDR) asymptote.
(At low enough $\mathrm{Re}$, laminarization was also observed after the flow passes the Virk asymptote, before another type of instability emerges~\citep{Choueiri_Hof_PRL2018}.)
Before MDR, distinction is further made more recently between low-extent (LDR) and high-extent drag reduction (HDR)~\citep{warholic1999influence}.
Starting from the Newtonian limit and with increasing $\mathrm{Wi}$, the flow undergoes a series of transitions between four different stages of behaviors: pre-onset, LDR, HDR, and MDR~\citep{Xi_Graham_JFM2010}.

LDR and HDR were first differentiated because their mean velocity profiles appear different in shape, which is observed in various experimental and numerical studies~\citep{warholic1999influence,ptasinski2003turbulent,li2006influence,mohammadtabar2017turbulent}.
Recall that Newtonian turbulent mean velocity profiles display the Prandtl-von K\'{a}rm\'{a}n (PvK) log law
\begin{gather}
	U^+=2.5\ln y^++5.5
	\label{Eq:PvK}
\end{gather}
across most of the near-wall layer ($y^+\gtrsim 30$)~\citep{Kim_Moin_JFM1987}. This log-law layer is connected to the near wall viscous sublayer via a buffer layer at $5\lesssim y^+\lesssim 30$~\citep{Pope_2000}.
At LDR, the buffer layer velocity profile raises up and its thickness also increases. Meanwhile the log-law layer stays parallel to the PvK log law only with a vertical offset (i.e., same slope but larger intercept compared with \cref{Eq:PvK} owing to the DR in the buffer layer).
At HDR, however, the slope of the mean velocity profile clearly increases in the log-law layer.
This effect was initially attributed to the quantitative magnitude of DR in earlier studies with $\mathrm{DR}\%\approx 35$ (
\begin{gather}
	\mathrm{DR}\%\equiv\frac{C_\text{f}-C_\text{f,s}}{C_\text{f}}\times100\%
\end{gather}
is the percentage drop of the friction factor $C_\text{f}$; subscript ``s'' indicates the solvent -- i.e., Newtonian benchmark fluid) often cited as the cutoff~\citep{warholic1999influence}.
Recent more systematic studies revealed that this transition is accompanied by a series of sharp changes in flow statistics and may occur at much lower $\mathrm{DR}\%$ at lower $\mathrm{Re}$~\citep{Xi_Graham_JFM2010,Zhu_Xi_JNNFM2018}.
Most notably, suppression of Reynolds shear stress (RSS) is mainly contained in the buffer layer at LDR which extends across the whole boundary layer at HDR.
In addition, the mean velocity profile was shown to no longer follow a logarithmic dependence at HDR~\citep{white2012re,Elbing_Perlin_PoF2013}.
All these evidences indicate that turbulent DR is a two-stage process with distinct mechanisms. The first is a localized weakening of turbulence concentrated in the buffer layer which starts at $\mathrm{Wi}_\text{onset}$. The second is a fundamental change in turbulent dynamics in the log-law layer that is only triggered at the LDR-HDR transition.
Fundamental understating of the second mechanism (HDR) is very limited which however has important implications in the area of flow control. In particular, existing non-additive based DR techniques mainly results in flow statistics characteristic of LDR~\citep{Deng_Xu_JTURB2016}.
Knowing how polymers trigger HDR will inspire new approaches that elevate the DR outcome to the next level.

Flow statistics and turbulent dynamics are often conceptualized in the framework of coherent structures such as vortices and streaks~\citep{robinson1991coherent, bernard1993vortex, adrian2007hairpin}.
These structures are commonly spotted in flow field images (from flow visualization experiments or direct numerical simulations -- DNS) and provides a vehicle for describing mechanisms of turbulent self-sustaining processes and momentum transport\RevisedText{~\citep{panton2001overview,jimenez2007we,wallace2016quadrant,jimenez2013near}}.
Attempts have also been made to establish the relationship between the mean velocity profile and the underlying coherent structures~\citep{lozano2012three}.
For instance, \citet{perry1995wall} attributed the logarithmic dependence (\cref{Eq:PvK}) to the population of highly lifted-up vortices. 
For viscoelastic turbulence, it is commonly accepted that polymer stresses can cause DR by suppressing the motion of vortices~\citep{DeAngelis_Piva_CompFl2002,dubief2005new,Li_Graham_JFM2006,kim2007effects,li2015simple}, which offers a convincing explanation for the onset of DR.
Much less is known about the second stage of DR as the LDR-HDR transition was not considered a qualitative change in turbulent dynamics until very recently~\citep{Zhu_Xi_JNNFM2018}.
Quantitative analysis of vortex distribution revealed that sharp changes in flow statistics coincide with the start of coherent structure localization, with HDR characterized by spotty clusters of vortices separated by laminar-like regions~\citep{Zhu_Xi_JNNFM2018}, which corroborates the earlier description of the intermittent transitions between active and hibernating turbulence~\citep{Xi_Graham_PRL2010,Xi_Graham_JFM2012,Xi_Graham_PRL2012,Xi_Bai_PRE2016}.
Based on this, \citet{Zhu_Xi_JNNFM2018} hypothesized that the LDR-HDR transition stems from a fundamental change in the turbulence regeneration mechanism and the two-stage DR process is a reflection of two different modes of polymer effects on turbulent structures.
At lower $\mathrm{Wi}$, polymers cause an across-the-board weakening of vortices and thus the onset of DR.
At higher $\mathrm{Wi}$ they start to suppress vortex lift-up and prevent its subsequent bursting events.
Since bursting can lead to the spreading of flow disturbances and trigger streak instability elsewhere in the domain~\citep{hamilton1995regeneration,schoppa2002coherent}, its suppression effectively blocks this pathway for vortex regeneration and exposes the more localized parent-offspring mechanism -- generation of new vortices at the edge of existing ones -- as the main process for turbulence sustenance at HDR.
Prevention of vortex lift-up also offers an explanation for the breaking of the mean velocity log law at HDR.

Like all studies of turbulent coherent structures, although there is no shortage of anecdotal evidences for this conceptual model, systematical analysis of changes in vortex configuration without subjective bias is a non-trivial challenge.
Conditional sampling has been an influential tool in the coherent structure analysis of viscoelastic turbulence, which averages the flow structures extracted based on events such as velocity ejection~\citep{kim2007effects,kim2008dynamics} and occurrence of streamwise vortices~\citep{sibilla2005near}.
Its outcome has significantly contributed to the fundamental understanding in this area, especially that of vortex suppression by polymer forces which causes the transition into the first DR stage at $\mathrm{Wi}_\text{onset}$ (as reviewed above).
However, focusing on the average smears the variation between individual vortex objects and loses the information on the statistical distribution. Reliance on the predetermined detection events also limits its representativeness when studying dynamics involving complex vortex topologies and motions.
Proper-orthogonal decomposition (or Karhunen-Lo{\`e}ve analysis) was also widely used~\citep{DeAngelis_Piva_PRE2003, Housiadas_Beris_POF2005, Wang_Graham_AIChEJ2014, mohammadtabar2017turbulent}, which is most effective for quantifying energy distribution between flow modes of different length scales but information on real individual vortices is still missing.
A method that can extract individual realizations of vortex objects and objectively analyze their configurations and topologies can contribute new insight especially to the second stage of DR which, as discussed above, may involve more complex vortex dynamics. 

At the conceptual level, this is achieved in a two-step process: (1) vortex identification -- determining which regions in the flow field display vortical motions -- and (2) tracking -- grouping these regions into individual vortex objects.
\RevisedText{For vortex identification, its necessity} may not be obvious at first sight as one would intuitively turn to the vorticity field $\mbf\omega\equiv\mbf\nabla\times\mbf v$ for describing swirling flows.
The limitation of vorticity becomes clear when we consider a simple shear flow where, despite the absence of any vortex, still has a vorticity magnitude proportional to the shear rate.
Most commonly used 
vortex identification criteria
are based on scalar identifiers calculated from the velocity gradient tensor $\mbf\nabla\mbf v$~\citep{jeong1997coherent,chong1990general,hunt1988eddies}.
Here, we \RevisedText{illustrate with} the $Q$-criterion~\citep{hunt1988eddies} \RevisedText{which is} used in this study.
For incompressible flow, the $Q$ quantity is defined as
\begin{equation} 
Q=\frac{1}{2}(\|\mbf\Omega\|^2-\|\mbf S\|^2), 
\label{equ_Q}
\end{equation}
where
$\Vert\cdot\Vert$ denotes the Frobenius tensor norm: e.g., $\Vert\mbf\Omega\Vert\equiv\sqrt{\sum_i\sum_j\Omega_{ij}^2}$. The strain-rate tensor, $\mbf{S}\equiv\left(\mbf{\nabla}\mbf{v}+\mbf{\nabla}\mbf{v}^T\right)/2$, and the vorticity tensor, $\mbf{\Omega}\equiv\left(\mbf{\nabla}\mbf{v}-\mbf{\nabla}\mbf{v}^T\right)/2$, are the symmetric and antisymmetric parts of $\mbf{\nabla}\mbf{v}$, respectively. 
\Cref{equ_Q}, on its face, can be interpreted as a comparison between the magnitudes of fluid rotation (measured by $\|\mbf\Omega\|^2$) and strain ($\|\mbf S\|^2$).
The magnitude of $Q$ provides a basis for categorizing flow regions based on their local kinematics.
Regions with large positive $Q$ are dominated by strong rotation and thus correspond to vortices.
Regions with large negative $Q$ are dominated by strain -- i.e., stretching of fluid elements, which indicates extensional flow.
For a strict shear flow, it is easily verifiable that $Q=0$.
The reader is referred to \citet{Xi_Bai_PRE2016} for a more quantitative discussion on the relationship between $Q$ and local flow type.
A similar argument was also adopted by the recent studies of \citet{pereira2017elliptical,pereira2017statistics} which divided viscoelastic flow fields into regions with different $Q$ magnitudes. Energy exchanges between these $Q$ regions were analyzed to understand polymer-turbulence dynamics.
\RevisedText{The $Q$-criterion is just one of many vortex identification criteria available in the literature~\cite{chong1990general,Jeong_Hussain_JFM1995,liu2018rortex,dong2019new}.
Another widely-known example is the $\lambda_2$-criterion proposed by \citet{Jeong_Hussain_JFM1995}, in which $\lambda_2$ is the second largest eigenvalue of the $\mbf{S}^2+\mbf{\Omega}^2$ tensor and flow regions with negative $\lambda_2$ (similar to the positive-$Q$ situation) are considered to be dominated by vortex motions. 
Comparison between different vortex identification criteria has been widely studied in the literature and it is generally agreed that in complex turbulent flow fields, results from most common criteria are by and large
}
equivalent\RevisedText{~\citep{chakraborty2005relationships,chen2015comparison}}.
A more detailed introduction of vortex identification was provided in our earlier paper~\citep{Zhu_Xi_JFM2019}.

Much less development was seen in vortex tracking. 
Scalar fields of the identifier, e.g., $Q$ \RevisedText{and $\lambda_2$}, can be easily visualized by rendering its three-dimensional isosurfaces, although care must be taken in the selection of the threshold level~\citep{chu1993direct,lozano2012three,lozano2014time}.
This makes vortex objects easy to identify by eyes but not by a computer program for quantitative analysis.
A vortex tracking algorithm will enable the identification of individual vortex objects and quantification of their location, size, and topology without the subjectivity of human intervention.
A classical example is the method of \citet{jeong1997coherent}, which identifies vortex axes -- center-lines around which the fluid rotates in a swirling motion -- by stitching together local planar maxima of the identifier. The extracted axis-lines can be used in conditional sampling studies to align individual vortex objects for averaging~\citep{jeong1997coherent,hussain1987eduction,Zhu_Xi_JPhysCS2018}.
This method was however designed only for (quasi-)streamwise vortices whose axis-lines extend in nearly-straight lines aligned with the mean flow.
These vortices are important for the self-sustaining process of turbulence at least at lower $\mathrm{Re}$~\citep{Waleffe_POF1997} and DR in the buffer layer~\citep{Li_Graham_JFM2006}: the latter, as reviewed above, is responsible for LDR.
Vortices of more complex configuration, such as hairpin vortices with $\Omega$-shaped axis-lines, are of broad interest to many outstanding areas of research, including turbulence regeneration at high $\mathrm{Re}$, dynamics in the log-law layer, and bypass transition to turbulence~\citep{adrian2007hairpin,Wu_Moin_PoF2009,Schlatter_Henningson_PoF2008}.
In the case of viscoelastic flow concerned here, complex three-dimensional vortices are key to the understanding of HDR.
Recall that HDR is marked by qualitative changes in the turbulent statistics of the log-law layer~\citep{Zhu_Xi_JNNFM2018} where highly curved vortices are expected to play a more important role.
The mechanism proposed in \citet{Zhu_Xi_JNNFM2018} for the LDR-HDR transition also requires the understanding of polymer effects on lifted-up vortices, which are again significantly curved away from the streamwise direction.

Motivated by these, \citet{Zhu_Xi_JFM2019} have recently developed a new method termed ``vortex tracking by iterative propagation'' or VATIP.
The method borrows the original idea of \citet{jeong1997coherent} of extracting vortex axis-lines by connecting points along their pathways and introduces an iterative search process to connect new points for axis-line propagation in all three spatial dimensions.
It has been shown to successfully capture vortices with more general three-dimensional configurations, including those with curved axis-lines, non-streamwise alignment, or complex branched topology.
A vortex classification procedure was also proposed in the same study which sorts vortices identified by VATIP into commonly-observed types, such as quasi-streamwise vortices, hooks, hairpins, and irregularly branched ones.

The development of VATIP has enabled for the first time statistical analysis of vortex distribution and conformations.
This study will leverage this new tool to investigate polymer effects on vortex dynamics in different stages of viscoelastic turbulence.
Although much attention has been dedicated to the vortex-polymer interaction in the literature, this is the first time that the statistical distribution of vortex configuration and topology can be quantitatively analyzed and compared between different $\mathrm{Wi}$ in an unbiased manner.
Special focus is on the LDR-HDR transition, where knowledge of the dynamics of complex hairpin-like vortices is particularly important, and how the changing vortex dynamics may be responsible for the observed changes in the mean flow.
As shown later, our results lead to extensive evidences for the lift-up suppression mechanism hypothesized in \citet{Zhu_Xi_JNNFM2018} and, perhaps more importantly, the first complete description of vortex dynamics that accounts for both LDR and HDR.
The paper is organized as follows. In \cref{Sec_method}, we will describe our simulation protocol and provide a brief introduction to the VATIP algorithm.
We will then start the results part in \cref{Sec_flo_stat} with flow statistics and highlight their changes between the LDR and HDR stages. This includes the quadrant analysis of velocity fluctuations as an indirect measurement of the changes in coherent structures.
Direct visualization of vortex configurations at different stages will be compared in \cref{Sec_ins_vor}, where the capability of VATIP in vortex tracking will also be demonstrated.
The extracted vortex axis-lines will then be statistically analyzed in \cref{Sec_vor_shape,Sec_vor_type}.
After polymer effects on different aspects of vortex dynamics are investigated, the paper will conclude with a physical description of the vortex dynamics behind the two DR stages (in \cref{Sec_conclude}).

\section{Formulation and methodology}\label{Sec_method}
\subsection{Direct numerical simulation}\label{Sec:method:DNS}

\begin{figure}
	\centering				
	\includegraphics[width=.98\linewidth, trim=0mm 0mm 0mm 0mm, clip]{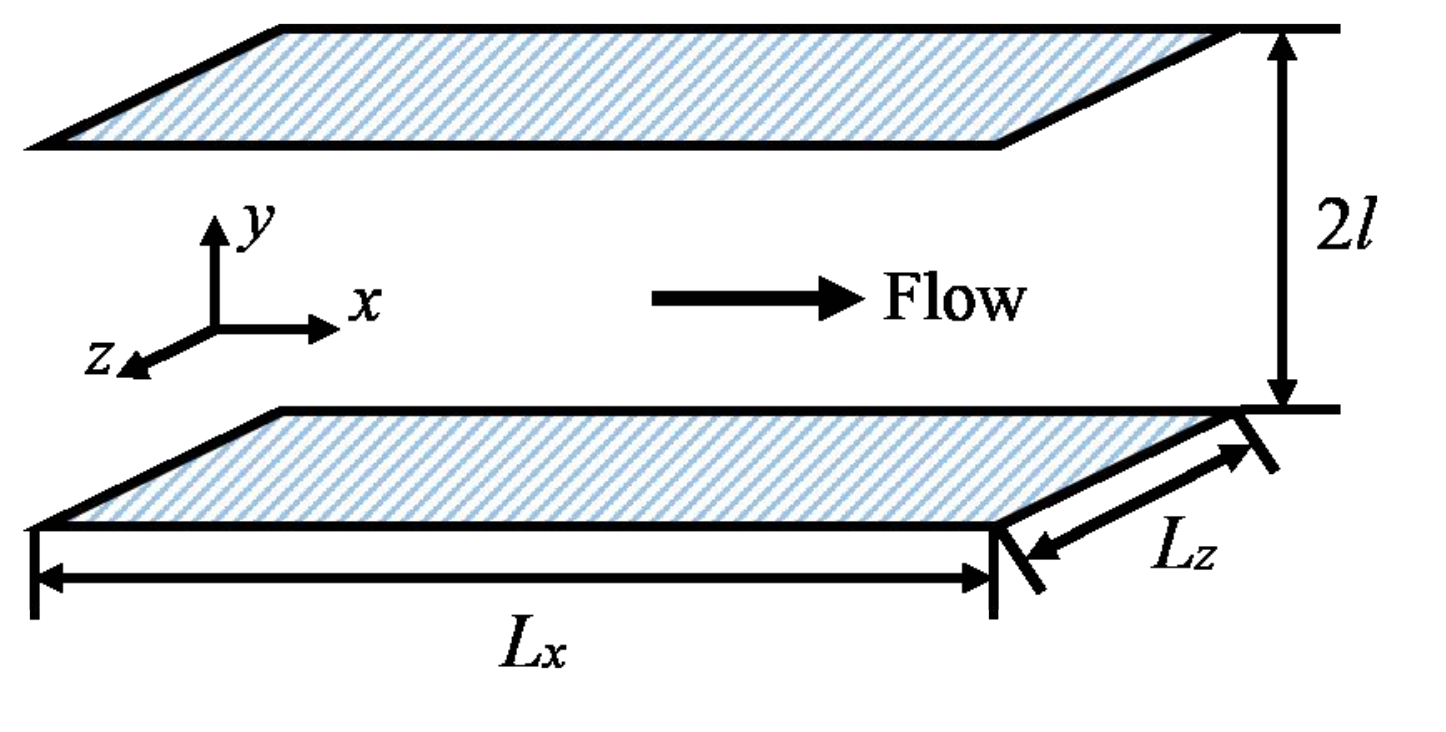}			
	\caption{Schematic of the flow geometry.}\label{fig:geomery}
\end{figure}

DNS in plane Poiseuille flow (the geometry is shown in \Cref{fig:geomery}) is implemented in this study. The flow is driven by a constant pressure drop and is oriented in the x-direction. 
The simulation domain size is $L_x\times 2l\times L_z$.
Variables in the simulation are nondimensionalized by the turbulent outer units. That is, lengths are normalized by the half-channel height $l$, velocities by the laminar centerline velocity $U_c$, pressure by $\rho U_c^2$ (where $\rho$ is the fluid density: i.e., for viscoelastic cases, it is the density of the polymer solution), and time by $l/U_c$.

Governing equations for the polymeric turbulence are summarized as
\begin{linenomath}\begin{equation}
	\frac{\partial\boldsymbol{v}}{\partial t}+\boldsymbol{v}\cdot \boldsymbol{\nabla}\boldsymbol{v}=-\boldsymbol{\nabla}p+\frac{\beta}{\mathrm{Re}}\nabla^2\boldsymbol{v}+\frac{2\left(1-\beta\right)}{\mathrm{Re}\mathrm{Wi}}\left(\boldsymbol{\nabla}\cdot\boldsymbol{\tau }_\text{p}\right),\label{equ_momentum}
	\end{equation}\end{linenomath}
\begin{linenomath}\begin{equation}
	\boldsymbol{\nabla}\cdot\boldsymbol{v}=0\label{equ_continuity}
	\end{equation}\end{linenomath}
and
\begin{linenomath}\begin{equation}
	\begin{split}
	\frac{\partial\boldsymbol{\alpha }}{\partial t}+\boldsymbol{v}\cdot\boldsymbol{\nabla}\boldsymbol{\alpha }-\boldsymbol{\alpha }\cdot\boldsymbol{\nabla} \boldsymbol{v}-\left(\boldsymbol{\alpha }\cdot\boldsymbol{\nabla} \boldsymbol{v}\right)^\mathrm{T}
	\\
	=\frac{2}{\mathrm{Wi}}(-\frac{\boldsymbol{\alpha }}{1-\frac{\mathrm{tr}\left(\boldsymbol{\alpha }\right)}{b}}+\frac{b\boldsymbol{\delta}}{b+2})+\frac{1}{\mathrm{ScRe}}\nabla^2 \boldsymbol{\alpha},
	\end{split}
	\label{equ_Fene1}
	\end{equation}\end{linenomath}
\begin{linenomath}\begin{equation}
	\boldsymbol{\tau}_\text{p}=\frac{b+5}{b}\left(\frac{\boldsymbol{\alpha}}{1-\frac{\mathrm{tr}\left(\boldsymbol{\alpha }\right)}{b}}-\left(1-\frac{2}{b+2}\right)\boldsymbol{\delta }\right).\label{equ_Fene2}
	\end{equation}\end{linenomath}
In \cref{equ_momentum}, the Reynolds number $\mathrm{Re}$ and corresponding friction Reynolds number $\mathrm{Re}_\tau$ are defined as $\mathrm{Re}\equiv\rho U_cl/\eta$ and $\mathrm{Re}_\tau\equiv\rho u_\tau l/\eta$ ($u_\tau$ is the friction velocity), respectively. The two Reynolds numbers can be directly related through $\mathrm{Re}_\tau=\sqrt{2\mathrm{Re}}$.
The Weissenberg number measures the level of elasticity and is defined as the product of the polymer relaxation time $\lambda$ and the mean wall shear rate, i.e., $\mathrm{Wi}\equiv 2\lambda U_c/l$.
The viscosity ratio $\beta\equiv\eta_\text{s}/(\eta_\text{s}+\eta_\text{p})$ is the ratio of the solvent viscosity to the total zero-shear-rate viscosity of the polymer solution (subscripts ``s'' and ``p'' indicate solvent and polymer contributions to viscosity, respectively). 
The contribution of polymers to the flow momentum is accounted for by the last term on the right-hand side (RHS) of \cref{equ_momentum}, where $\mbf\tau_\text{p}$ is the polymer stress tensor.
The FENE-P constitutive equations (\cref{equ_Fene1,equ_Fene2})~\citep{bird1987dynamics}, where polymer molecules are treated as finitely extensible nonlinear elastic (FENE) dumbbells, are adopted in this study to calculate $\mbf\tau_\text{p}$. In FENE-P, $\boldsymbol{\alpha}$ represents the polymer conformation tensor and is defined as $\boldsymbol{\alpha}\equiv\langle\boldsymbol{Q}\boldsymbol{Q} \rangle$ , where $\boldsymbol{Q}$ denotes the end-to-end vector of the dumbbell.
The maximum extensibility parameter $b$ constrains the length of polymer dumbbells through $\max(\mathrm{tr}(\boldsymbol{\alpha}))\leq b$.
The last term on the RHS of \cref{equ_Fene1} $(1/(\mathrm{ScRe}))\nabla^2 \boldsymbol{\alpha}$ ($\mathrm{Sc}$ is the Schmidt number) is an artificial diffusion \RevisedText{(AD)} term (not part of the FENE-P model) introduced for the sole purpose of maintaining numerical stability. The use of \RevisedText{AD} is required for the DNS of viscoelastic fluid flows using pseudo-spectral methods (see below). The practice is well studied and established in the literature~\citep{Sureshkumar_Beris_JNNFM1995}.

The Poiseuille flow implies periodic boundary conditions in the x- (streamwise) and z- (spanwise) directions, meaning that all variables are continuous across domain boundaries: e.g., $\mbf v(L_x,y,z)=\mbf v(0,y,z)$.
In the y- (wall-normal) direction, the no-slip boundary condition is applied to the parallel walls for the velocity field: i.e.,
\begin{gather}
	\mbf v=0\quad\text{at}\;y=\pm l.
\end{gather}
The original FENE-P equation does not require boundary conditions in the $y$-direction. Adding \RevisedText{AD} introduces second-order partial derivatives and changes the mathematical nature of the equation, for which wall boundary conditions are now required.
\RevisedText{We follow the standard procedure originally proposed by \citet{Sureshkumar_Beris_JNNFM1995} (and widely used by researchers~\citep{pereira2017elliptical,Housiadas_Beris_POF2005,Ptasinski_Nieuwstadt_JFM2003,Li_Khomami_PRE2015,lopez2018dynamics,Xi_Graham_JFM2010}), in which the b}oundary values of $\mbf\alpha$ are computed at each time step by directly integrating \cref{equ_Fene1} in time for grid points at the walls ($y=\pm l$ or $\pm1$ after nondimensionalization) without the \RevisedText{AD} term.
These values then provide boundary conditions for solving the equation, including AD, for the rest of the channel.
\RevisedText{The rationale behind this treatment is that the AD term is not part of the physical model and by solving the equation without AD, the solution is at least strictly accurate at the boundaries. (For the rest of the channel, a small AD is necessary for numerical stability.)
Detailed implementation of this boundary treatment is provided in \cref{App_A}.}

\begin{table}
	\begin{center}
		\begin{tabular}{c|cccc|cccc|cc}
			\hline
			$\mathrm{Re}_\tau$	&$\mathrm{Wi}$	&$\beta$	&$b$	&$\mathrm{Sc}$
			&$\delta_x^+$	&$\delta_z^+$	&$N_y$	&$\delta_t$
			&$\mathrm{DR\%}$	&Stage\\
			\hline
			$172.31$	&vary	&$0.97$	&$5000$ &$0.3$
			&$9.09$	&$5.44$	&$195$	&$0.01$
			&vary	&vary\\	
			\hline
			\multirow{2}{*}{400}
			&25	&$0.9$	&$900$	&$0.25$
			&\multirow{2}{*}{$9.09$}	&\multirow{2}{*}{$5.44$}	&\multirow{2}{*}{$473$}	&\multirow{2}{*}{$0.005$}
			&16.8			&LDR	\\
				&50	&$0.9$	&$3600$	&$0.25$
			&&&&
			&41.2			&HDR	\\
			\hline
		\end{tabular}
		\caption{Physical parameters and numerical settings of viscoelastic DNS simulations.}
		\label{tab:DNSparam}
	\end{center}	
\end{table}

DNS results of two different $\mathrm{Re}$ are analyzed with VATIP in this study.
The lower $\mathrm{Re}$ case, i.e., $\mathrm{Re}=14845$ ($\mathrm{Re}_\tau=172.31$), uses the same dataset previously reported in \citet{Zhu_Xi_JNNFM2018}.
At this $\mathrm{Re}$, a clear transition between LDR and HDR is already clearly observable with all features of the transition captured.
Also, for Newtonian flow, this $\mathrm{Re}$ is sufficient to produce a pronounced PvK log-law layer~\citep{Zhu_Xi_JFM2019}.
Simulation runs at a wide range of $\mathrm{Wi}$ with fixed $\beta$ and $b$ (see \cref{tab:DNSparam}) have been performed at this $\mathrm{Re}$, including multiple cases in both LDR and HDR stages.
At the higher $\mathrm{Re}=80000$ ($\mathrm{Re}_\tau=400$), two viscoelastic cases are simulated. The parameters are so selected that one is at LDR and the other at HDR.
Newtonian flow is also simulated for both $\mathrm{Re}$. Parameters for the DNS runs reported in this study are summarized in \cref{tab:DNSparam}.

A Fourier-Chebyshev-Fourier pseudo-spectral scheme is adopted to discretize all variables in space. The spatial periods are $L_x^+\times L_z^+=4000\times 800$
for all simulations at both $\mathrm{Re}$. (The superscript ``$+$'' represents quantities nondimensionalized with inner scales -- velocities by $u_\tau$ and lengths by $\eta/\rho u_\tau$).
An $N_x\times N_z=440\times 147$ mesh is used for the $x$ and $z$ Fourier transforms and
Chebyshev-Gauss-Lobatto points are used for the Chebyshev transform in the y-direction. 
The number of grid points $N_y$ is adjusted with $\mathrm{Re}$ (see \cref{tab:DNSparam}): for $\mathrm{Re}=172.31$, the range of $y$-grid spacing $\delta_y^+$ is \numrange{0.022}{2.79} (minimum at the walls and maximum at the channel center) and for $\mathrm{Re}_\tau=400$, it is \numrange{0.011}{3.03}.
The time integration chooses a third-order semi-implicit backward-differentiation/Adams-Bashforth scheme\RevisedText{ (BDAB3)}~\citep{peyret2002spectral}. 
Different time step sizes are chosen at the two $\mathrm{Re}$ (\cref{tab:DNSparam}) according to the Courant-Friedrichs-Lewy (CFL) stability condition.
The magnitude of the numerical diffusivity $1/(\mathrm{ScRe})$ (in the \RevisedText{AD} term of \cref{equ_Fene1}) is \num{2.25e-4} for $\mathrm{Re}=172.31$ and \num{5e-5} for $\mathrm{Re}=400$, respectively. This
is lower than most studies in the literature in which a numerical diffusivity in the order of $O(0.01)$ is generally found to be safe~\citep{ptasinski2003turbulent, sureshkumar1997direct, Sureshkumar_Beris_JNNFM1995, Dimitropoulos_Beris_JNNFM1998}. 
A detailed numerical sensitivity analysis at three different levels of numerical diffusivity and resolution was reported in \citet{Zhu_Xi_JNNFM2018} and not repeated here.
The viscoelastic DNS code used in this study is custom-developed by expanding the open-source package for Newtonian DNS \texttt{ChannelFlow}, originally developed by John~F.~Gibson~\citep{gibson2012channelflow} and later improved and parallelized by Tobias Schneider, Hecke Degering (Schrobsdorff), and co-workers~\citep{Tuckerman_Schneider_PoF2014}.

\subsection{VATIP for vortex tracking}

\begin{figure}
	\centering
	\includegraphics[width=.98\linewidth, trim=1mm 1mm 1mm 1mm, clip]{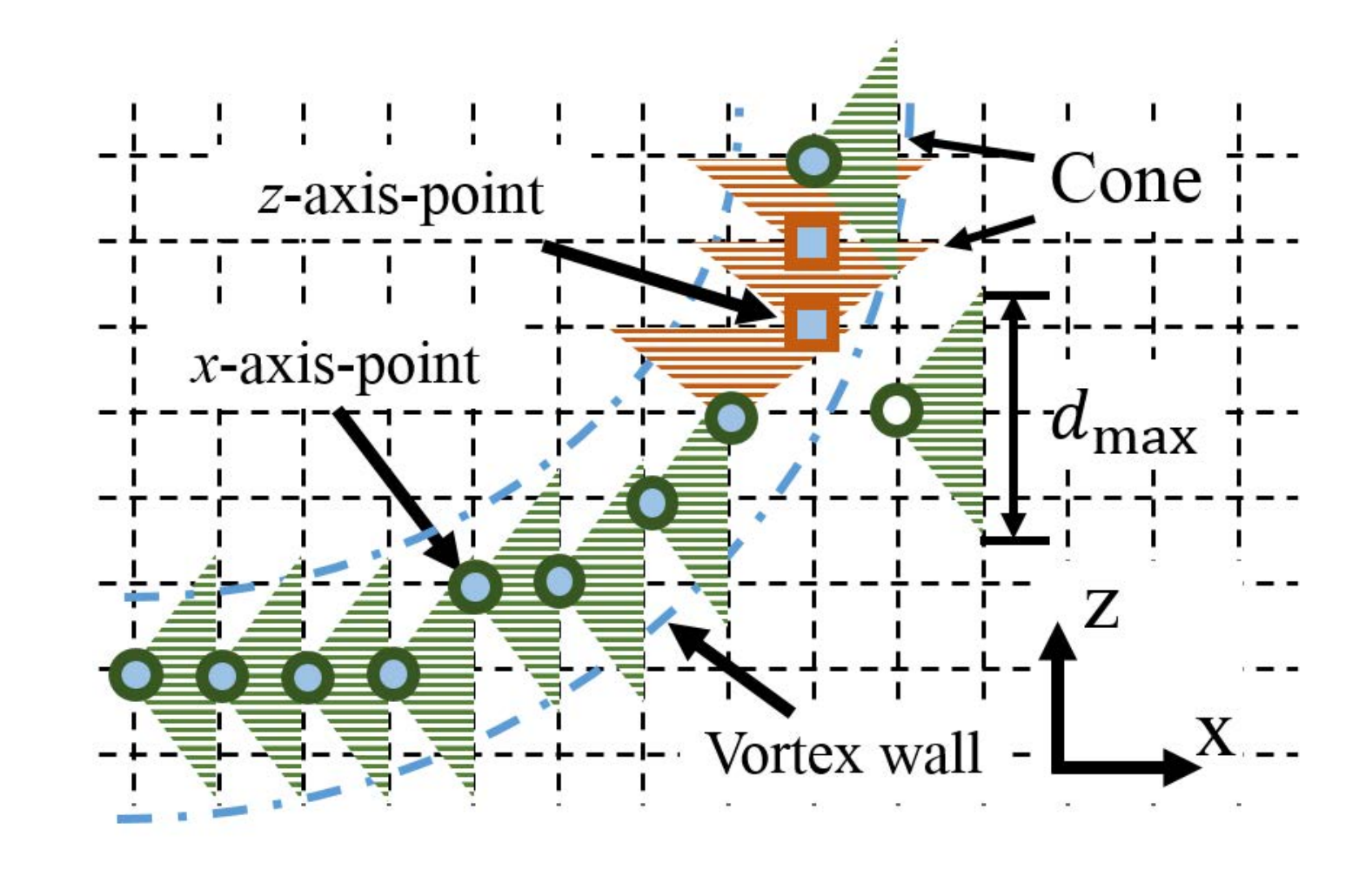}
	\caption{The conceptual plot of the VATIP algorithm. A new point is connected to a propagating axis-line if it falls within a detection cone. The $x$-direction search round looks for local maxima of $Q$ in the $yz$ plane (labeled $x$-axis-points); the search continues in other directions after no more $x$-axis-points can be added. For simplicity, the plot only sketches a two-dimensional scenario without explicitly showing the search round in the $y$-direction. 
	The triangles thus represent the planar projection of the detection cones.}
	\label{fig:track_method}
\end{figure}

\begin{table}
	\begin{center}
		\begin{tabular}{c|ccc}
			\hline
			$\mathrm{Re}_\tau=172.31$	&Newt.	&$\mathrm{Wi}=20$	&$\mathrm{Wi}=80$	\\
			\hline
			$Q_\text{rms}$	&$0.0325$	&$0.017$	&$0.0061$	\\
			\hline
			$\mathrm{Re}_\tau=400$	&Newt.	&LDR	&HDR	\\
			\hline
			$Q_\text{rms}$	&$0.0305$	&$0.0125$	&$0.00461$	\\
			\hline
		\end{tabular}
		\caption{Values of $Q_\text{rms}$ in representative Newtonian and viscoelastic DNS flow fields.}
		\label{tab:Qrms}
	\end{center}	
\end{table}

The purpose of VATIP is to extract the axis-lines of individual vortices around which the fluid rotates. 
If a vortex is defined as a tube in which the scalar identifier $Q$ exceeds a curtain threshold, the $Q$ magnitude increases from the tube shell inwards and peaks at the axis. The axis-line preserves the position, size, shape, and topology of the vortex and is thus particularly instrumental in vortex analysis.
The scalar $Q$ field first needs to be calculated from the velocity data (\cref{equ_Q}). 
To determine the threshold value of $Q$ for vortex identification, we follow a systematic procedure based on the so-called ``percolation analysis'', which has been extensively discussed in previous studies~\citep{Zhu_Xi_JNNFM2018,Zhu_Xi_JFM2019}.
In short, a very low $Q$ threshold will over-identify vortex regions and render one interconnected (percolating) vortex structure whereas at the other limit (high threshold), vortices will be under-identified with many valid vortices excluded from the result.
The percolation analysis identifies $Q$ values at which individual vortex objects are just separated apart but are still mostly preserved.
In this study, spatial regions with $Q>0.4Q_\text{rms}$ ($Q_\text{rms}$ being the root-mean-square -- RMS -- value of $Q$ over the domain
\begin{equation}
Q_\text{rms}\equiv\sqrt{\frac{1}{2lL_xL_z\Delta T}\int_0^{\Delta T}\left(\iiint Q^2 dxdydz\right)dt}
\label{Eq:Qrms}
\end{equation}	
) are identified as vortex regions.
Values of $Q_\text{rms}$ for several representative cases (in different flow stages) are provided in \cref{tab:Qrms}.
Notably, $Q_\text{rms}$ decreases monotonically with increasing $\mathrm{DR\%}$, indicating the correlation between vortex weakening and drag reduction. More detailed results and discussion in this regard are found in our earlier study~\citep{Zhu_Xi_JNNFM2018}.

Each point on the axis-line is the maximum of $Q$ in the corresponding cross-sectional plane of the vortex tube. Depending on the direction of the vortex segment concerned, the axis-point may appear as a local two-dimensional maximum in the  $yz$, $xz$, or $xy$ plane (for vortex segments aligned in the $x$, $y$, or $z$ direction, respectively).
Therefore, all two-dimensional local maxima in planes of all orientations within the identified vortex regions need to be found and recorded as potential axis-points.

Connecting axis-points that belong to the same vortices to form axis-lines is the central task of vortex tracking which is illustrated in \cref{fig:track_method}.
The process starts with $yz$ planes for $x$-direction tracking. At each $yz$ grid plane, a new axis-line is initiated from each unassociated potential axis-point. Existing axis-lines attempt to propagate along the $x$ direction by finding eligible axis-points in the next plane for connection.
Connection is made if the next axis-point falls within a cone-shaped region projected from the propagating end of the axis-line. The size of the cone is determined from the average radius of a streamwise vortex tube
\begin{gather}
	r_\text{v}=\sqrt{\frac{\sum_{i=1}^{N_x}A_{\text{v},i}}{\pi \sum_{i=1}^{N_x} N_{\text{v},i}}},
	\label{Eq:estiRadus}
\end{gather}
(where $N_x$ is the number of $x$-grid points -- i.e., the number of $yz$-planes, $i$ is the $yz$-plane index, $A_{\text{v},i}$ is the total area of vortex regions on plane $i$ 
calculated by adding up all areas that satisfy the vortex identification criterion ($Q>0.4Q_\text{rms}$ in this study) on the plane, and $N_{\text{v},i}$ is the number of separate vortex areas on the plane) and a base diameter of $d_\text{max}=1.4r_\text{v}$ is used in this study.
This so-called ``cone-detective'' idea was first proposed by \citet{jeong1997coherent} which however only focused on streamwise vortices and their algorithm stops the search after the $x$-direction search round.
In VATIP, the search continues in the $y$ and then $z$ direction for vortices whose axis-lines are no longer confined in the $x$-direction.
These continued search rounds extend the existing axis-lines in new directions by connecting axis-points in two-dimensional planes of other orientations: e.g., for the search in the $z$-direction, local $Q$ maxima in $xy$-planes, which are termed $z$-axis-points, are added to the growing axis-lines when they fall into the detection cones (now pointed towards the $z$ direction; see \cref{fig:track_method}).
Initiation of new axis-lines is not allowed in these continued search rounds to avoid false identification (i.e., all axis-lines are initiated in the first round of $x$-direction search).
However, separate axis-lines are allowed to merge if the detection cone from the propagating end finds another axis-line within its range.
Consider a hairpin vortex typically observed in the log-law layer~\citep{robinson1991coherent,adrian2007hairpin,Wu_Moin_PoF2009} with an $\Omega$-shaped axis configuration, its two legs extend towards the wall and along the $x$-direction and will be captured with the first $x$-direction search round; at the downstream end, the legs lift up away from the wall (which requires $y$-direction search) and merge along the $z$-direction to form an arc (which requires $z$-direction search and axis-line merging).
An $x$-$y$-$z$ search cycle would successfully capture such vortices. Many vortices observed in DNS results, however, do not conform to this canonical shape and in order to capture a wider variety of three-dimensional vortices with complex axis-line topology, the VATIP algorithm continues to iteratively loop over searches in all three directions until the number of identified vortices converges.

\RevisedText{We use the $Q$-criterion in our studies, but the VATIP algorithm can be easily adapted to any other scalar vortex identifier as long as it maps quantitatively to the intensity of vortical flow. For example, in the case of the $\lambda_2$-criterion, one only needs to replace $Q$ in the above procedure with $-\lambda_2$. (Minus sign is added because $\lambda_2<0$ indicates vortices and is thus equivalent to $Q>0$.)}

VATIP was tested with intentionally generated curved vortices such as hooks and hairpins as well as actual DNS flow fields. It was shown to successfully capture vortices of all known shapes and configurations typically observed in near-wall turbulence~\citep{Zhu_Xi_JNNFM2018}.
Note that this section only provides a high-level description of the key elements of VATIP. The readers are referred to \citet{Zhu_Xi_JFM2019} for implementation details and further discussions about the method.

\section{Results and Discussion}\label{Sec_results}
\subsection{Flow statistics}\label{Sec_flo_stat}

\begin{figure}
	\centering
	\includegraphics[width=.8\linewidth, trim=5mm 0mm 8mm 0mm, clip]{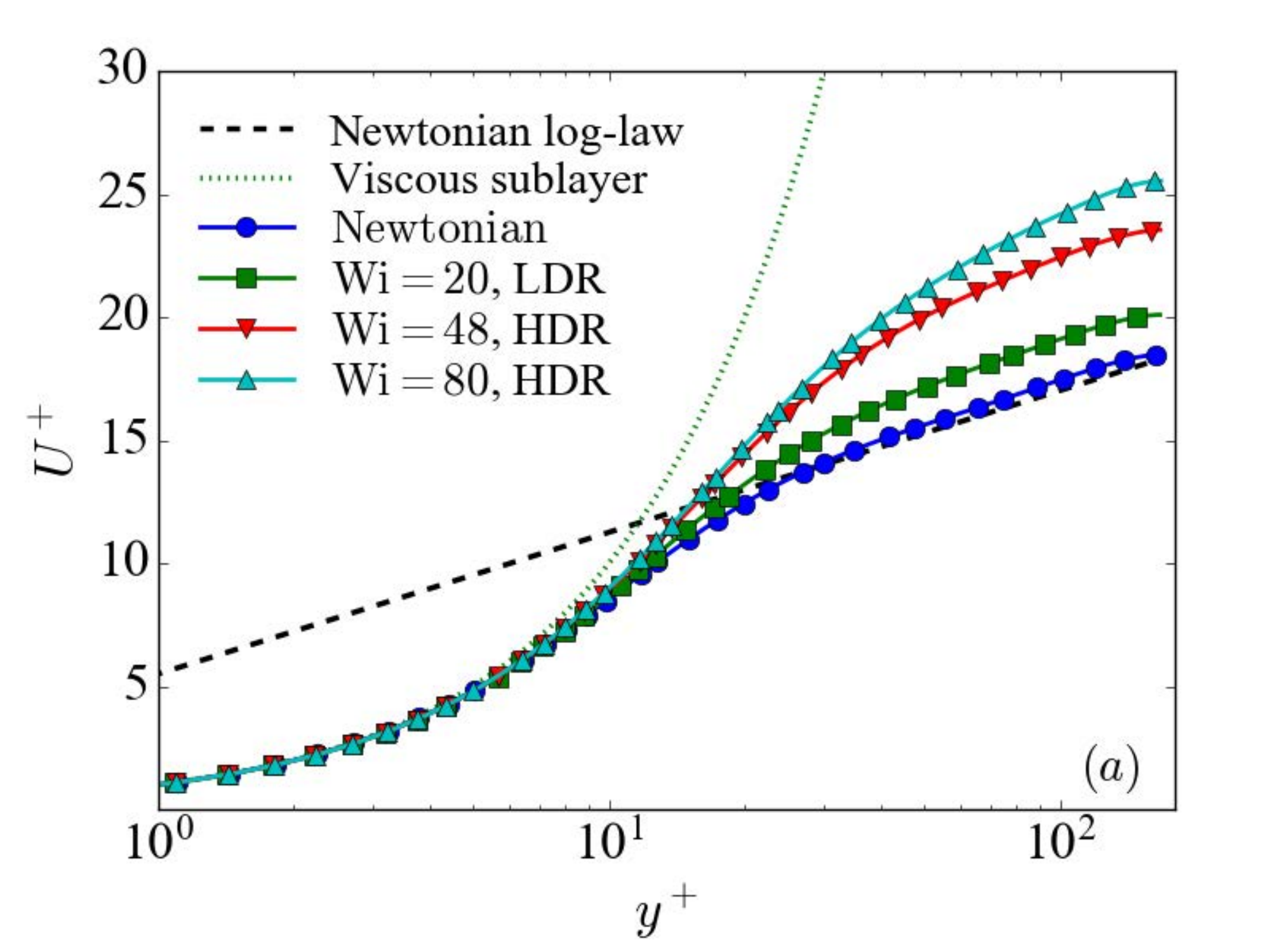}
	\\
	\includegraphics[width=.8\linewidth, trim=5mm 0mm 8mm 0mm, clip]{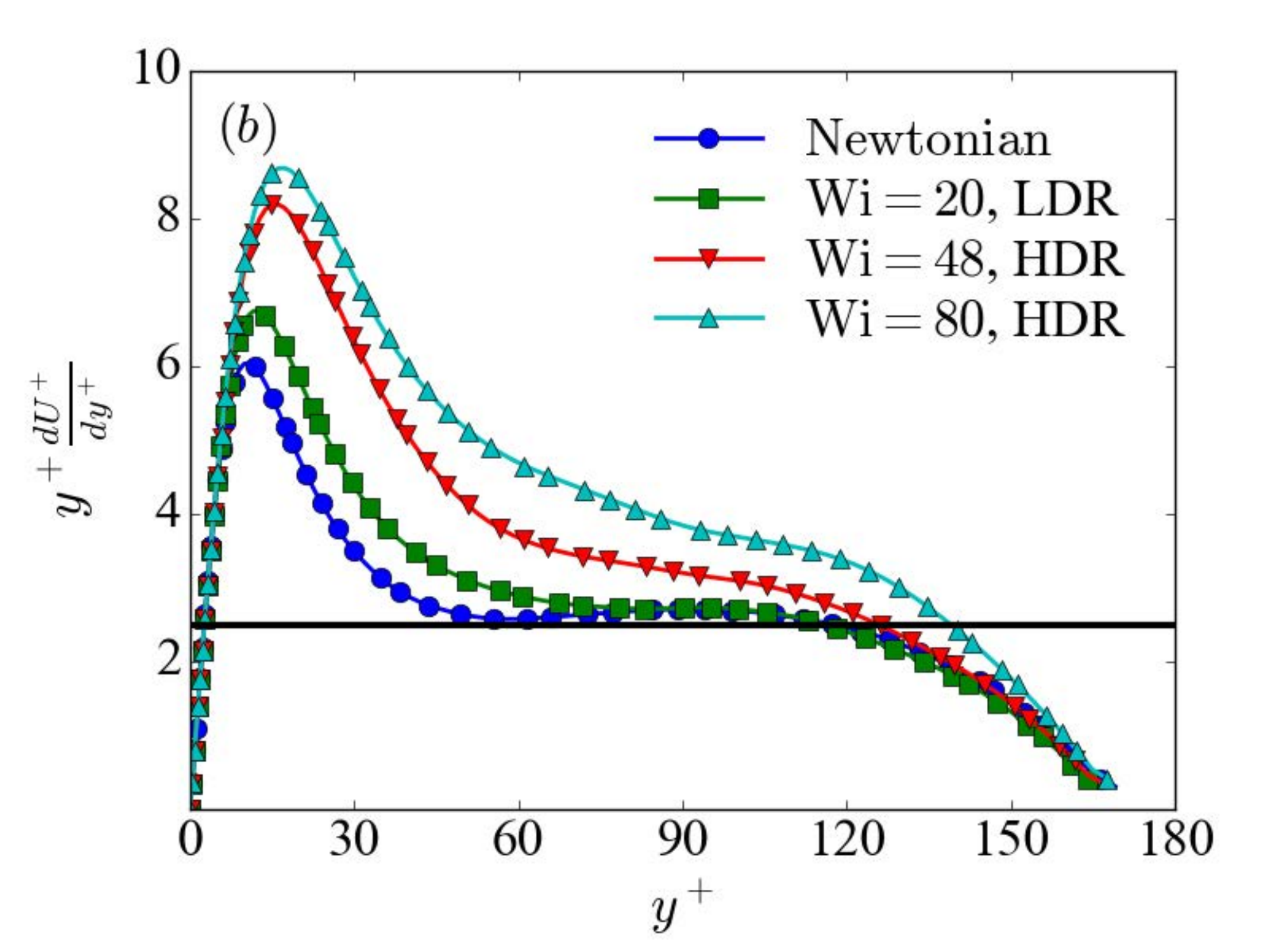}
	\caption{(\textit{a}) Mean velocity profiles ($U^+$ vesus $y^+$) and (\textit{b}) log-law indicator function ($y^+dU^+/dy^+$ vesus $y^+$) at $\mathrm{Re}_\tau=172.31$; horizontal line marks the PvK magnitude of 2.5 (\cref{Eq:PvK}).}
	\label{fig:mean_vel} 
\end{figure}

By injecting polymers into turbulent flows, properties of the flows are significantly changed which leads to considerable reduction of the friction drag and increase of the mean flow rate.
In \cref{fig:mean_vel}(a), we show the mean velocity profiles of the Newtonian and three viscoelastic cases ($\mathrm{Wi}=20,48,$ and $80$) at $\mathrm{Re}_\tau=172.31$.
For the Newtonian case, the profile closely follows the PvK asymptote (\cref{Eq:PvK}) at $y^+ \gtrsim 50$, indicating that the log law layer has been sufficiently developed at this $\mathrm{Re}_\tau=172.31$.
For the $\mathrm{Wi}=20$ case, the velocity profile lifts up in the buffer layer ($20\lesssim y^+\lesssim 50$) but stays parallel to the PvK asymptote at higher $y^+$. By contrast, the profiles of the $\mathrm{Wi}=48$ and $80$ cases lift up across most of the channel including what used to be the log-law layer.
This observation has been the most-discussed difference between LDR and HDR in the literature~\citep{warholic1999influence, ptasinski2003turbulent,li2006influence,Housiadas_Beris_POF2003,Xi_Graham_JFM2010,Zhu_Xi_JNNFM2018,mohammadtabar2017turbulent}.
In our case, it is clear that $\mathrm{Wi}=20$ belongs to LDR and $\mathrm{Wi}=48$ and $80$ are well within the HDR regime.
The qualitative change in the mean velocity gradient is more clearly seen in the 
logarithmic law indicator function (\cref{fig:mean_vel}(b)).
Note that any $U^+(y^+)$ dependence can be written in the generic form of 
\begin{gather}
	U^+=\frac{1}{\kappa}\ln y^++B
	\label{Eq:loglawgen}
\end{gather}
where $B$ is a constant and the indicator function
\begin{gather}
	\frac{1}{\kappa}=\frac{dU^+}{d\ln y^+}=y^+\frac{dU^+}{dy^+}
	\label{eq:indicator}
\end{gather}
is a constant only if the profile follows a logarithmic dependence.
For Newtonian and LDR ($\mathrm{Wi}=20$) cases, a clear inflection point with $1/\kappa\approx 2.5$ shows up at $y^+ \approx 50$, which is followed by a nearly flat segment at $50\lesssim y^+\lesssim 100$ -- a clear log-law layer.
For HDR cases ($\mathrm{Wi}=48$ and $80$), the inflection point disappears and the segment at larger $y^+$ is no longer flat.
This indicates the log law is no longer valid at the HDR stage, which is consistent with the finding of \citet{white2012re}.

\begin{figure}
	\centering
	\includegraphics[width=.98\linewidth, trim=0mm 0mm 0mm 0mm, clip]{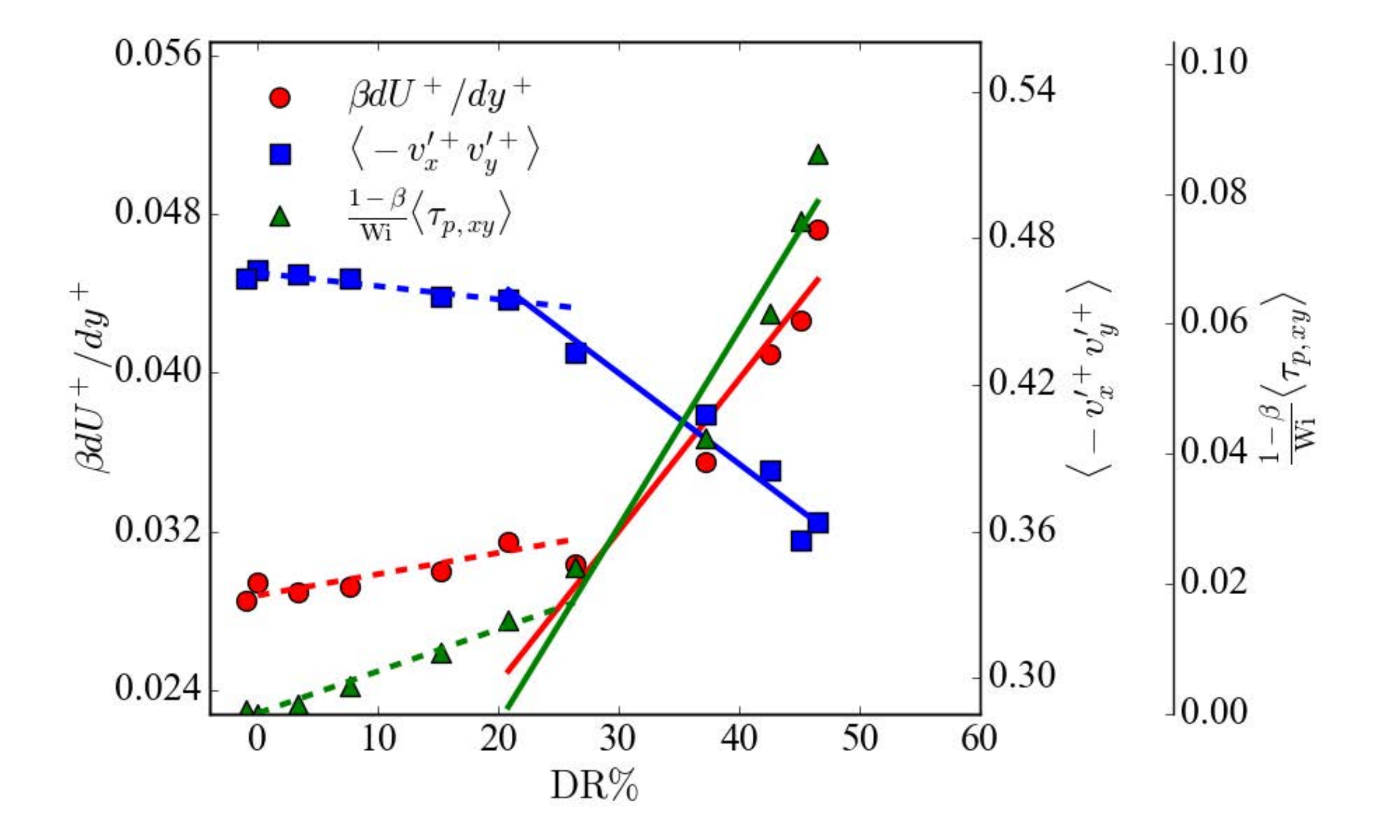}
	\caption{Shear stress components at $y^+=103.2$ plotted against $\mathrm{DR\%}$ ($\mathrm{Re}_\tau=172.31$), including the Newtonian case ($\mathrm{DR\%}$) and viscoelastic cases at $\mathrm{Wi} = 8$, 12, 16, 20, 24, 32, 48, 64, 80, and 96  ($\mathrm{DR\%}$ increases monotonically with $\mathrm{Wi}$ with the exception of $\mathrm{Wi}=8$, which is pre-onset and nearly overlaps with the Newtonian case). The lines are guides to the eyes for the LDR (dashed) and HDR (solid) stages.}
	\label{fig:SSB_y06} 
\end{figure}

The mean velocity gradient (which determines the indicator function -- \cref{eq:indicator}) is related with velocity fluctuation and polymer stress through the shear stress balance: 
\begin{equation}
\langle\tau_{xy}^{+}\rangle =\beta \frac{dU^+}{dy^+}+\langle -v_x^{\prime +}v_y^{\prime +} \rangle +\frac{1-\beta}{\mathrm{Wi}}\langle\tau_{\text{p},xy}\rangle
\label{equ_stress_bal}
\end{equation}
where the three terms on the RHS represents contributions from the viscous, Reynolds, and polymer shear stresses, respectively
($\langle\cdot\rangle$ represents averages over $x$, $z$, and $t$ axes).
Under constant mean pressure gradient, the total shear stress is a constant for given $\mathrm{Re}$ and $y^+$ position --
\begin{gather}
	\langle\tau_{xy}^{+}\rangle=1-\frac{y^+}{\mathrm{Re}_\tau}.
\end{gather}
With increasing $\mathrm{DR}\%$, the rise of viscous and polymer shear stresses must be accompanied by the drop of RSS.
Recent studies further showed that, similar to the change of $1/\kappa$, the suppression of RSS is contained within and near the buffer layer at LDR and significant reduction of RSS at larger $y^+$ is only obvious at HDR~\citep{Xi_Graham_JFM2010,Zhu_Xi_JNNFM2018}.
In \cref{fig:SSB_y06}, the magnitudes of these shear stress components at $y^+=103.2$ (which is well within the log-law layer for the Newtonian case) are plotted against $\mathrm{DR\%}$ for DNS results at $\mathrm{Re}_\tau=172.31$, including the Newtonian and viscoelastic cases at ten different $\mathrm{Wi}$ (see caption of \cref{fig:SSB_y06}).
The LDR-HDR transition occurs at $\mathrm{DR\%}\approx20\% $ and $\mathrm{Wi}\approx24$, which is marked by a sharp turn in all three components.
Variations in these quantities are mild at LDR but for HDR their $\mathrm{DR\%}$-dependencies become steep.
The rapid decline of RSS, in particular, indicates a new stage of turbulence suppression in the log-law layer which is only initiated at the start of HDR.
Note that the transition point of $\mathrm{DR}\%\approx 20\%$ is not universal and at higher $\mathrm{Re}$ the critical $\mathrm{DR}\%$ will be higher. Although earlier studies widely quoted $\mathrm{DR}\%\approx 30\sim35\%$ as the separation between LDR and HDR~\citep{warholic1999influence,li2006influence,li2015simple}, it was recently established that the transition point is a function of $\mathrm{Re}$~\citep{Zhu_Xi_JNNFM2018}, which again shows that the LDR-HDR transition is more than a quantitative effect of the level of $\mathrm{DR\%}$ but a shift between two qualitatively different stages of DR.

\begin{figure}
	\centering
	\includegraphics[width=.99\linewidth, trim=0mm 0mm 0mm 0mm, clip]{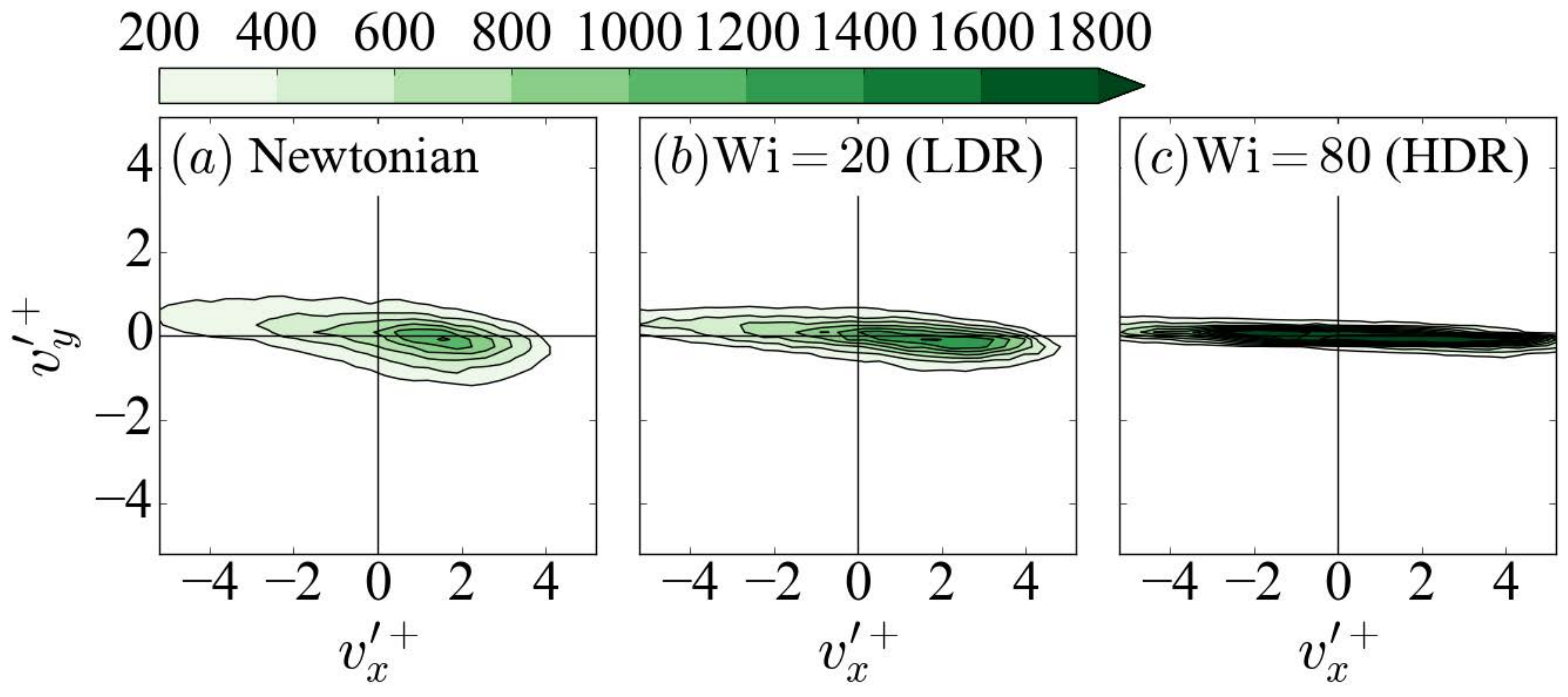}
	\caption{\label{fig:RSS_phase_yp25} Joint PDF of the streamwise and wall-normal velocity fluctuations at $y^+=25$ ($\mathrm{Re}_\tau=172.31$).}
\end{figure}

\begin{figure}
	\centering
	\includegraphics[width=.99\linewidth, trim=0mm 0mm 0mm 0mm, clip]{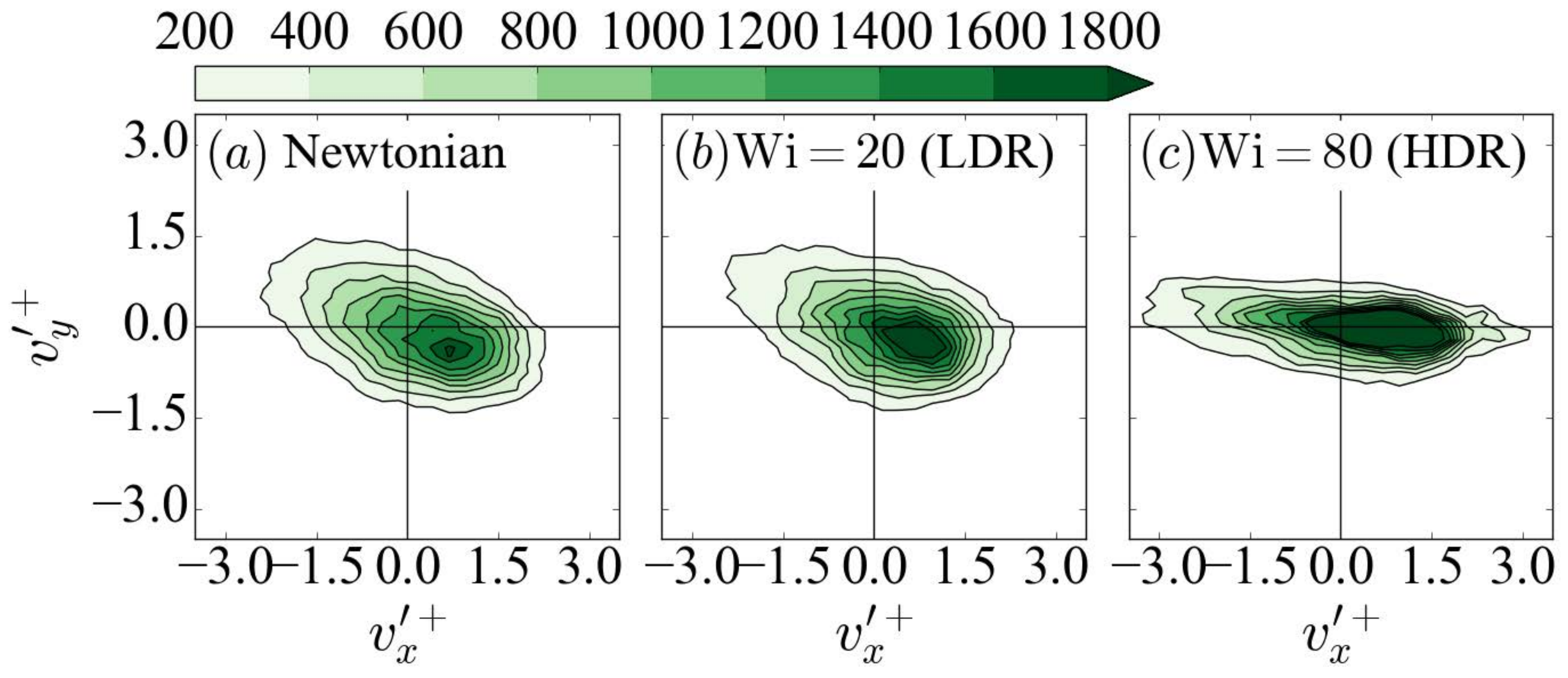}
	\caption{\label{fig:RSS_phase_yp100} Joint PDF of the streamwise and wall-normal velocity fluctuations at $y^+=100$ ($\mathrm{Re}_\tau=172.31$).}
\end{figure}

Velocity fluctuations at $\mathrm{Re}_\tau=172.31$ are inspected with quadrant analysis which plots the joint probability density function (PDF) between the streamwise and wall-normal velocity fluctuations (\cref{fig:RSS_phase_yp25,fig:RSS_phase_yp100}).
The distribution is typically skewed towards the second and fourth quadrants (Q2 and Q4) where $v_x^{\prime +}$ and $v_y^{\prime +}$ have opposite signs and thus contribute positively to the RSS (second term on the RHS of \cref{equ_stress_bal}).
The Q2 events, in which $v_x^{\prime +}<0$ and $v_y^{\prime +}>0$, correspond to the upward movement of the slower fluids near the wall to larger $y^+$ which causes a local reduction in the streamwise velocity and is often termed ``ejections''.
Meanwhile, the opposite Q4 events are called ``sweeps''~\citep{lozano2012three,wallace2016quadrant}.
The buffer layer (\Cref{fig:RSS_phase_yp25}) distribution is flatter owing to its stronger streamwise velocity fluctuations. 
As $\mathrm{Wi}$ increases, the joint PDF contour span shrinks in the $y$-direction while expands along $x$-direction, which is consistent with the established observation in the literature that the wall-normal and spanwise velocity fluctuations are suppressed by polymers but the streamwise fluctuations are often enhanced~\citep{sureshkumar1997direct, ptasinski2003turbulent, Min_Choi_JFM2003b, li2006influence}.
Suppression of the ejections and sweeps in the buffer layer reduces the wall-normal momentum fluxes responsible for the high Reynolds stress~\citep{townsend1980structure, marusic2010wall}.
Note that in the buffer layer, the joint PDF shape is already clearly modified in LDR, which only continues into HDR.
By contract, at higher $y^+$ (\cref{fig:RSS_phase_yp100}), the transition between LDR and HDR is sharp. The joint PDF patterns are similar between Newtonian and LDR cases whereas at HDR it is clearly flattened, indicating that polymer-induced changes in coherent motions only start at HDR in that wall region.
Our quadrant analysis results are remarkably similar to the recent experimental measurement by \citet{mohammadtabar2017turbulent} at comparable or lower $\mathrm{Re}$ ($\mathrm{Re}_\tau$ ranges from approximately 200 to 70 from the Newtonian limit to the highest DR\%).

Observations in flow statistics suggest that the LDR-HDR transition is underpinned by a sudden shift of the regions or wall layers where polymer interaction with turbulence is substantial.
At LDR, polymers mainly suppress turbulence in the buffer layer, causing its enlargement and higher mean velocity gradient, whereas the log-law layer is left largely intact. This is indeed the essence of the elastic sublayer theory of \citet{Virk_AIChEJ1975}.
The theory, however, does not account for the occurrence of the second stage of DR -- HDR -- where polymer effects on turbulent dynamics begin to substantially alter the log-law layer.
Further evidences for the transitions in flow statistics, as well as the localization in turbulence distribution at HDR, are found in \citet{Zhu_Xi_JNNFM2018} and not repeated here.
The primary focus of this study is to investigate the changes in coherent structure dynamics behind these observations.

\begin{figure}
	\centering
	\includegraphics[width=.99\linewidth, trim=0mm 0mm 0mm 0mm, clip]{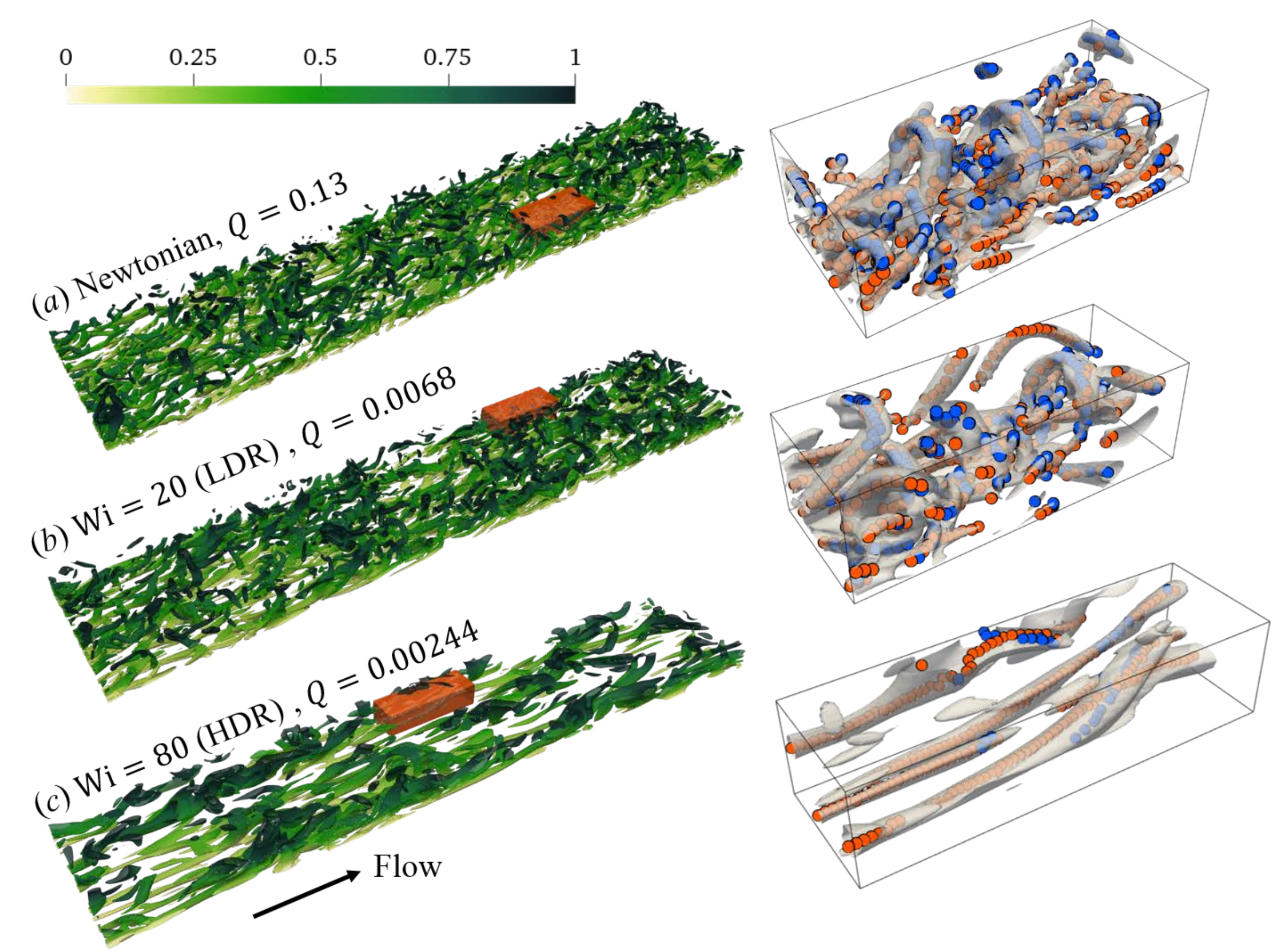}
	\caption{\label{fig:ins_vor} Instantaneous vortex structures of (\textit{a}) Newtonian, (\textit{b}) $\mathrm{Wi}=20$ and (\textit{c}) $\mathrm{Wi}=80$ cases at $\mathrm{Re}_\tau=172.31$ identified by the $Q$-criterion (only the bottom half of the channel is shown). The color shade (from light to dark) maps to the distance from the bottom wall in outer units.
	Part of the domain (orange box) is enlarged and shown on the right. Circular markers are axis-points identified by VATIP: orange (light) for x-axis-points; blue (dark) for y- and z-axis-points.}
\end{figure}

\subsection{Vortex conformation and tracking in instantaneous flow fields}\label{Sec_ins_vor}
We start with instantaneous images of flow-field vortices and their axis-line conformations identified by VATIP at $\mathrm{Re}_\tau=172.31$. Vortices are identified with the $Q$ criterion and the isosurfaces of $Q=0.4Q_\text{rms}$ are plotted in \cref{fig:ins_vor}. 
Although streamwise aligned vortices are seen in all cases especially near the wall, the Newtonian and LDR cases show strong tendency for vortex lift up, in which the vortex legs (in the upstream) are initiated near the wall along the streamwise direction but its head becomes detached from the wall in the downstream.
Detached vortex segments become distorted and deviate away from the flow direction.
Hairpins are a distinct type of lifted-up vortices with an $\Omega$-shaped contour: a transverse arc at the downstream end with two streamwise legs extending upstream towards the wall.
At this $\mathrm{Re}$, they are already populating the flow domain in the Newtonian and LDR cases.
The HDR image appear drastically different with significantly reduced instances of vortex lift-up, hairpins, and curved vortices. The vortices are more likely to stick close to the wall and become much more extended in the flow direction than LDR. This observation is consistent with the earlier observations in conditional eddies by \citet{kim2007effects}.
This dominance of elongated vortex conformation underlines the common observations of much smoother velocity distribution at HDR with extended streak patterns~\citep{Warholic_Hanratty_EXPFL2001, White_Mungal_EXPFL2004, Housiadas_Beris_POF2005, li2006influence}.

VATIP allows us to go beyond direct intuitive visual inspection and extract vortex conformations without subjective bias.
Vortex axis-points identified by VATIP are shown in \cref{fig:ins_vor} with circular markers for a smaller region in the domain.
It is clear that in all cases, the axis-lines (connecting all axis-points) obtained by VATIP successfully capture all visible vortices and well preserve their size, position, shape, and topology, including both straight (quasi-streamwise) and curved (e.g., hairpins) vortices.
Quasi-streamwise vortex axis-lines are mainly composed of $x$-axis-points (which are local maxima of $Q$ in $yz$-planes), represented by orange markers. For significantly lifted-up vortices (including hairpins), mostly seen in the Newtonian and LDR cases, $y$- and $z$-axis-points (blue markers; local maxima in $xz$- and $xy$-planes) must be included. This is a major improvement of VATIP compared with earlier approaches which are limited to streamwise vortices~\citep{jeong1997coherent,sibilla2005near}.
These $y$- and $z$- axis-points become less important at HDR where streamwise vortices dominate.
Spatial proximity between vortices in the DNS of full steady-state turbulence makes it difficult to clearly visualize individual vortex conformations.
More isolated vortices of a variety of shapes can be generated using transient DNS to test the VATIP performance, which was done in \citet{Zhu_Xi_JFM2019} and not repeated here.

\subsection{Polymer effects on vortex conformation and lift-up}\label{Sec_vor_shape}

\RevisedText{Recall that \citet{Zhu_Xi_JNNFM2018} hypothesized a change of vortex regeneration mechanism to explain the LDR-HDR transition. At LDR, similar to Newtonian flow, vortices can often be sustained and regenerated by streak instability.
Specifically, streamwise vortices lift up to higher wall-normal layers (such as the log-law layer), which gives rise to complex three-dimensional vortices such as hairpins. Lifted vortices tend to break up and burst into intense turbulent fluctuations, which can spread across the flow domain and trigger the instability of streaks elsewhere to generate more streamwise vortices.
(Despite the same self-sustaining mechanism as Newtonian turbulence, DR still occurs at LDR because of the overall weakening of vortices.)
At HDR, this vortex regeneration pathway is suppressed. Streamwise vortices stay confined closed to the wall (without lifting up) and become elongated by the flow. Instabilities of the shear layers between these vortices and the wall can generate new vortices near their ends, which is thus dubbed the ``parental-offspring'' mechanism.
In a later study~\citep{Zhu_Xi_JNNFM2019}, direct evidences were found for the bursting of vortices following lift up and the effects of polymers causing its suppression.}

\begin{figure}
	\centering
	\includegraphics[width=.99\linewidth, trim=0mm 0mm 0mm 0mm, clip]{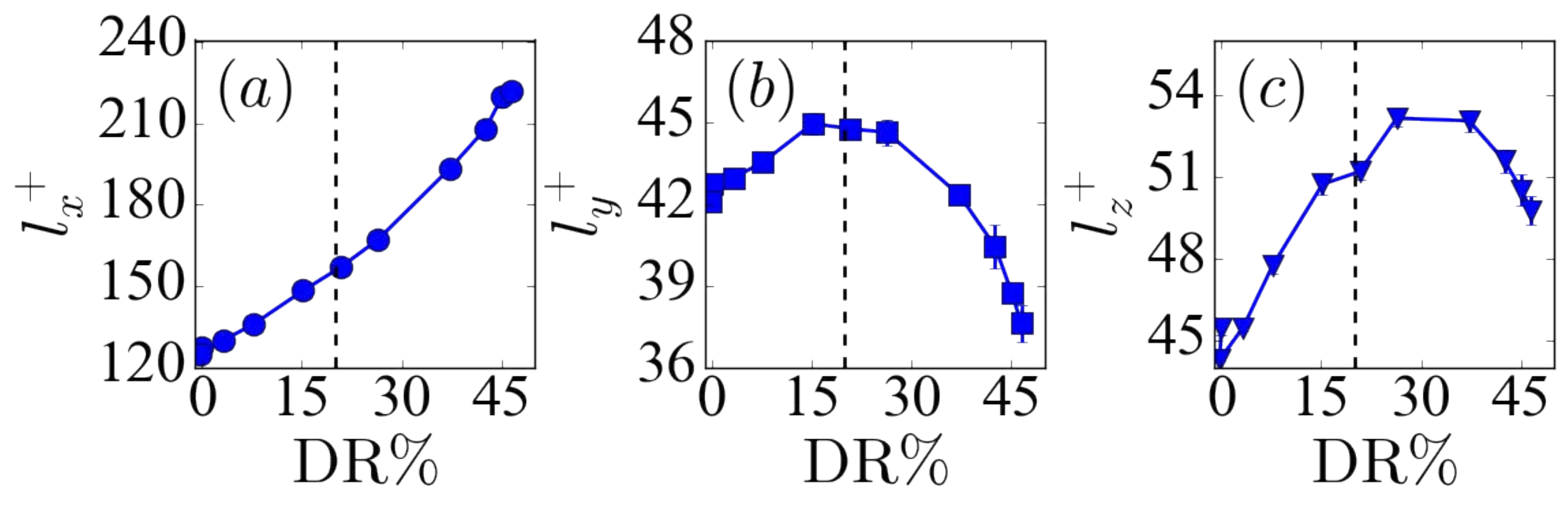}
	\caption{Average dimensions of the enclosing cuboid of each vortex at $\mathrm{Re}_\tau=172.31$: (\textit{a}) streamwise length $l_x^+$, (\textit{b}) wall-normal length $l_y^+$ and (\textit{c}) wall-normal length $l_z^+$. Dashed line marks the LDR-HDR transition.}
	\label{fig:vor_len}
\end{figure}

\RevisedText{From this hypothesis, one would expect the vortex conformation to statistically change at the LDR-HDR transition: e.g., lifted vortices would be less prevalent at HDR.}
Axis-lines extracted by VATIP open up the possibility for the statistical analysis of vortex conformations\RevisedText{, which offers the opportunity to directly test this picture}. \Cref{fig:vor_len} shows the average dimensions of vortices at $\mathrm{Re}_\tau=172.31$, measured by the edge lengths of a minimal cuboid enclosing each vortex.
A dashed line is drawn at $\mathrm{DR\%}=20\%$, which was identified earlier as the point of LDR-HDR transition at $\mathrm{Re}_\tau=172.31$ based on flow statistics (\cref{fig:SSB_y06}).
This line is provided in all $\mathrm{DR\%}$-dependence plots in this study to provide a reference for identifying the correlation, or the lack thereof (if that is the case), between changes in flow statistics and vortex structure measurements.
The average streamwise dimension of a vortex $l_x^+$ (\cref{fig:vor_len}(a)) increases nearly monotonically with $\mathrm{DR\%}$, indicating that vortices become elongated in the streamwise direction with polymer DR effects. This is indeed a well-established observation in the literature~\citep{li2006influence, li2015simple, kim2007effects, kim2008dynamics} and consistent with the direct observation in \cref{fig:ins_vor}.
The trend continues after the LDR-HDR transition with no notable change in pattern.
Streamwise vortex elongation can be interpreted as the result of vortex stabilization~\citep{dubief2005new,Zhu_Xi_JNNFM2019}: when a vortex does not lift up away from the wall or burst into pieces for an extended period of time, it is continuously stretched by the flow.
Because vortices are in general not strictly aligned with the $x$ axis, its elongation can also lead to increasing dimensions in the other directions $l_y^+$ and $l_z^+$.
This effect seems to dominate at LDR where $l_y^+$ and $l_z^+$ grow nearly monotonically (\cref{fig:vor_len}).
Due to the increasing stability of vortices, the wall-normal and spanwise length also increase in the LDR stage.
However, this trend is turned around
after the LDR-HDR transition. In \cref{fig:vor_len}(\textit{b}), the wall-normal length $l_y^+$ immediately drops when the HDR stage is reached, which is consistent with the hypothesis of \citet{Zhu_Xi_JNNFM2018} that at HDR polymers suppress the lift up of vortices.
Lift-up exposes the downstream end, or the ``head'', of the vortex to transverse flows, which bend the vortex sideways to form spanwise segments of vortex tubes (such as the arc in an $\Omega$-shaped hairpin vortex) and increase its dimension in $z$ direction $l_z^+$.
Suppression of vortex lift-up explains the reduction of curved vortices such as hairpins, as seen in \cref{fig:ins_vor}.
The spanwise vortex length $l_z^+$ (\cref{fig:vor_len}(\textit{c})) does indeed drop substantially at HDR. The turning point is slightly delayed compared with the LDR-HDR transition.
This seems to suggest that the start of HDR is more directly linked to lift-up suppression, which blocks the transfer of turbulent motions from the buffer layer to the log-law region, and the reduction of hairpins and spanwise vortex dimension is a secondary effect.
Highly lifted vortices will eventually burst into intense fluctuations~\citep{Zhu_Xi_JNNFM2019} which may seed new streak stabilities and lead to turbulence proliferation.
Confining the stabilized vortices to the streamwise direction leads to their prolonged stretching and a shift in the turbulent regeneration dynamics.

\begin{figure}
	\centering
	\includegraphics[width=.99\linewidth, trim=0mm 0mm 0mm 0mm, clip]{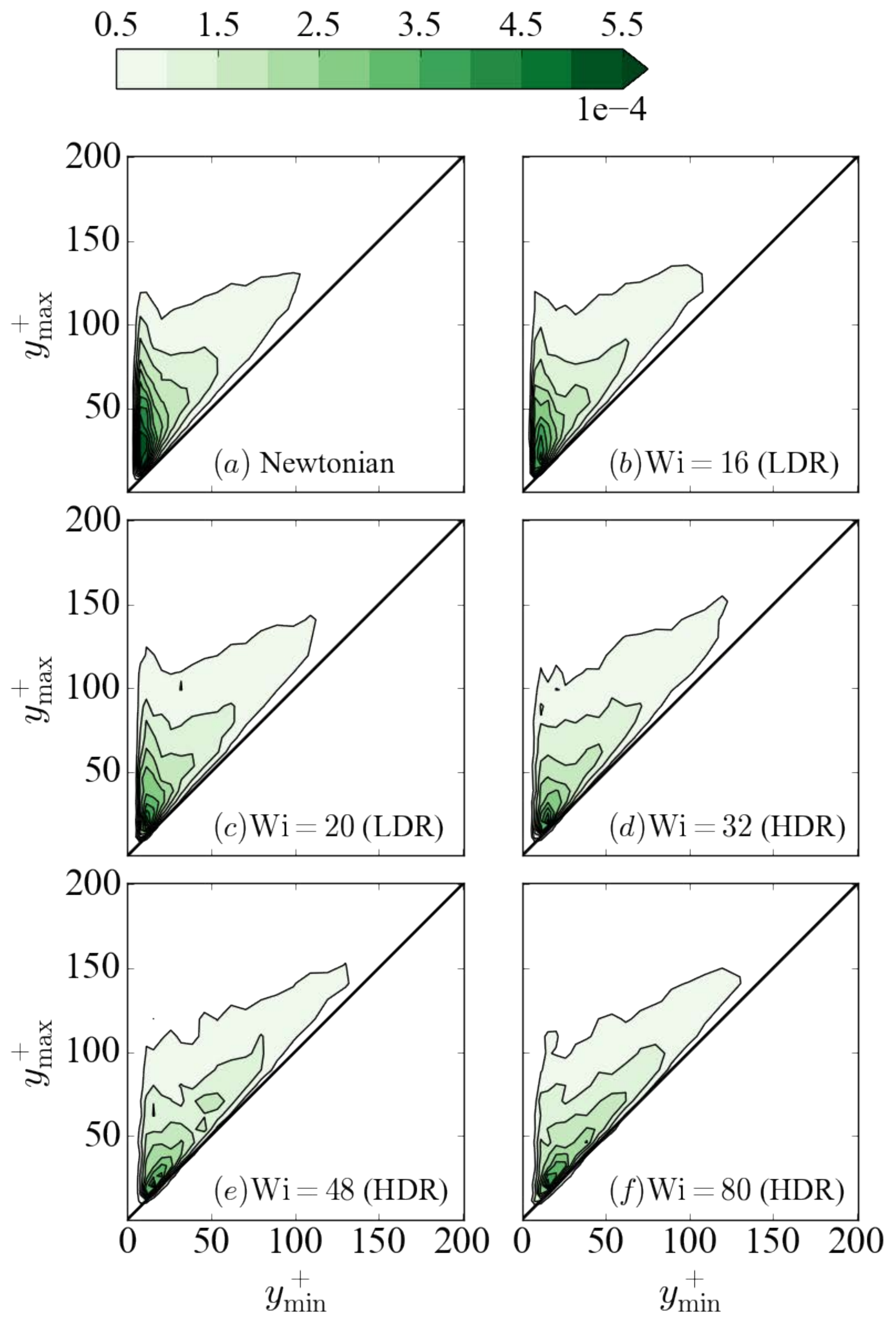}
	\caption{Joint PDFs of the wall-normal positions of the head and tail/legs of each vortex, as measured respectively by the maximum and minimum $y^+$ coordinates of the vortex axis-line, at $\mathrm{Re}_\tau=172.31$ and} different $\mathrm{Wi}$.
	\label{fig:vor_height_PDF}
\end{figure}

\begin{figure}
	\centering
	\includegraphics[width=.99\linewidth, trim=0mm 0mm 0mm 0mm, clip]{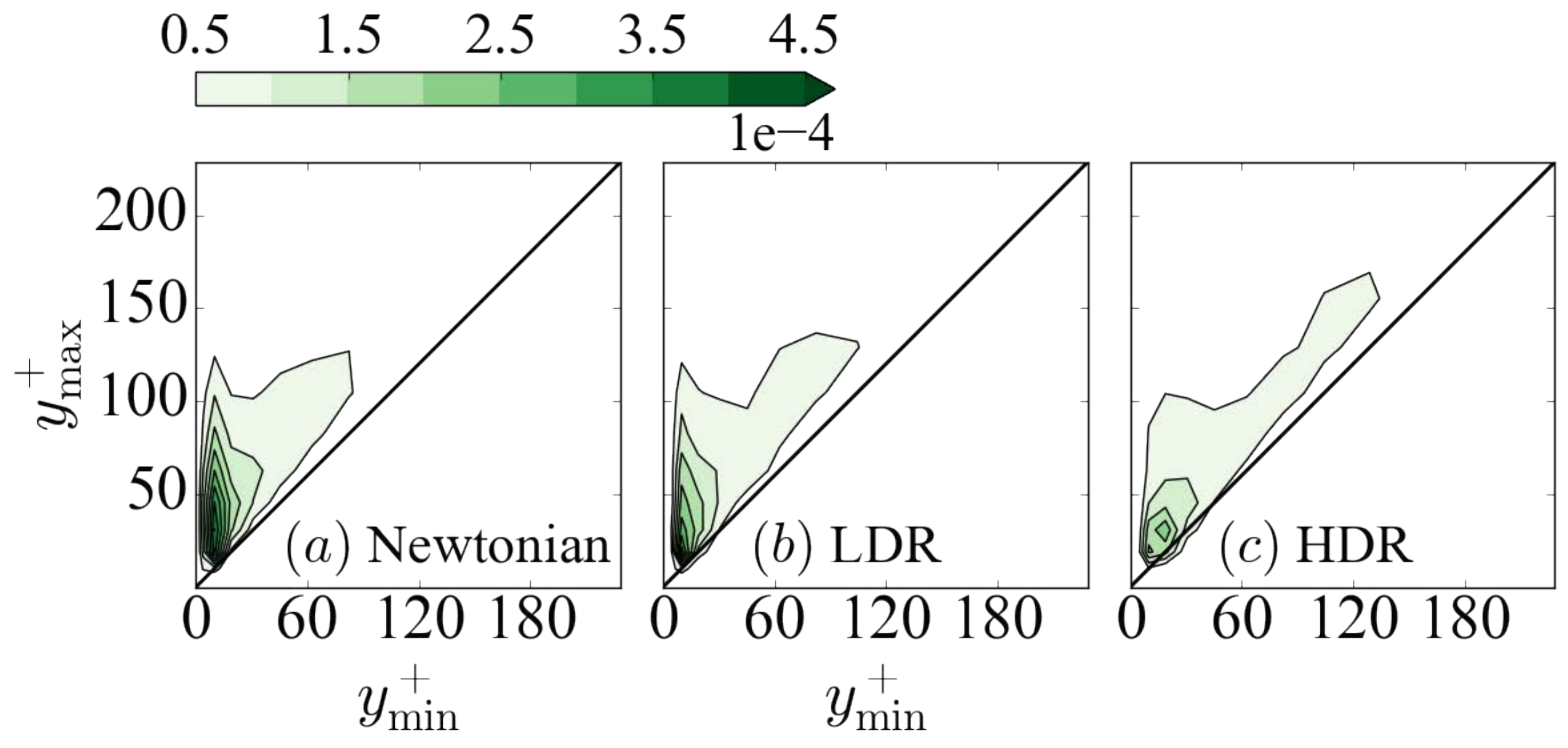}
	\caption{Joint PDFs of the wall-normal positions of the head and tail/legs of each vortex, as measured respectively by the maximum and minimum $y^+$ coordinates of the vortex axis-line, at $\mathrm{Re}_\tau=400$ and} different $\mathrm{Wi}$.
	\label{fig:vor_height_PDF_re80000}
\end{figure}

Vortex lift-up can now be quantified by the wall-normal positions of the head (highest point, typically at the downstream end) and leg(s)/tail (lowest point, typically at the upstream end) of the vortices. These positions can be measured from the axis-lines obtained from VATIP and the joint PDFs between them are shown in \cref{fig:vor_height_PDF}
for the $\mathrm{Re}=172.31$ case.
The distribution at LDR ($\mathrm{Wi}=16$ and $20$ cases) closely resembles that of the Newtonian case and is highly concentrated in the buffer layer ($y^+<30$). Two concentration bands extend from the peak distribution there: one along the vertical axis that corresponds to the highly lifted-up vortices (leg/tail $y^+_\text{min}$ in the buffer layer but head $y^+_\text{min}$ well into the log-law layer) and the other, slightly less populated, along the diagonal that corresponds to flat-lying vortices that align mostly along the streamwise direction.
The pattern clearly changes at HDR where the vertical band becomes significantly weakened and the diagonal band is more pronounced and extends to higher $y^+$. The concentration peak is still found in the buffer layer but it is now more aligned with the diagonal than the ordinate.
Distribution pattern at $\mathrm{Re}_\tau=400$ (\cref{fig:vor_height_PDF_re80000}) is strikingly similar, not only qualitatively (i.e., the pivot towards the diagonal) but also quantitatively. Wall-normal positions and spans of vortices are well comparable, in inner units, between these two distinctly different $\mathrm{Re}$, indicating strong scalability of coherent structures at different DR stages with increasing $\mathrm{Re}$.

From these results, it becomes clear that at LDR, despite an overall weakening of all vortices, vortex distribution has changed little compared with the Newtonian limit, whereas the suppression of vortex lift-up only starts at HDR, which corroborates our earlier notation that the LDR-HDR transition is a reflection of a new stage of DR with a distinct mechanism.
Earlier studies have suggested the possibility of lift-up suppression by polymers through direct flow field inspection or conditional sampling of average eddies~\citep{kim2007effects, Zhu_Xi_JPhysCS2018, Zhu_Xi_JNNFM2019}.
Statistical quantification of vortex lift-up tendency would not have been possible without the specific information on individual vortex axis-lines. 
More importantly, this is the first time polymer-induced lift-up suppression is associated with the LDR-HDR transition by direct evidence.
Vortex lift-up is important in the turbulent momentum transfer between different wall layers and widely believed to be responsible for the PvK log law (\cref{Eq:PvK})~\citep{townsend1980structure,perry1995wall,lozano2012three}. Its suppression at HDR thus offers a clear pathway to explain the changing mean flow profile in that regime.
Meanwhile, extension of the diagonal band indicates the increasing frequency of flat-lying vortices at higher wall layers, which again supports a change in the log-law dynamics.

\begin{figure}
	\centering
	\includegraphics[width=.99\linewidth, trim=0mm 0mm 0mm 0mm, clip]{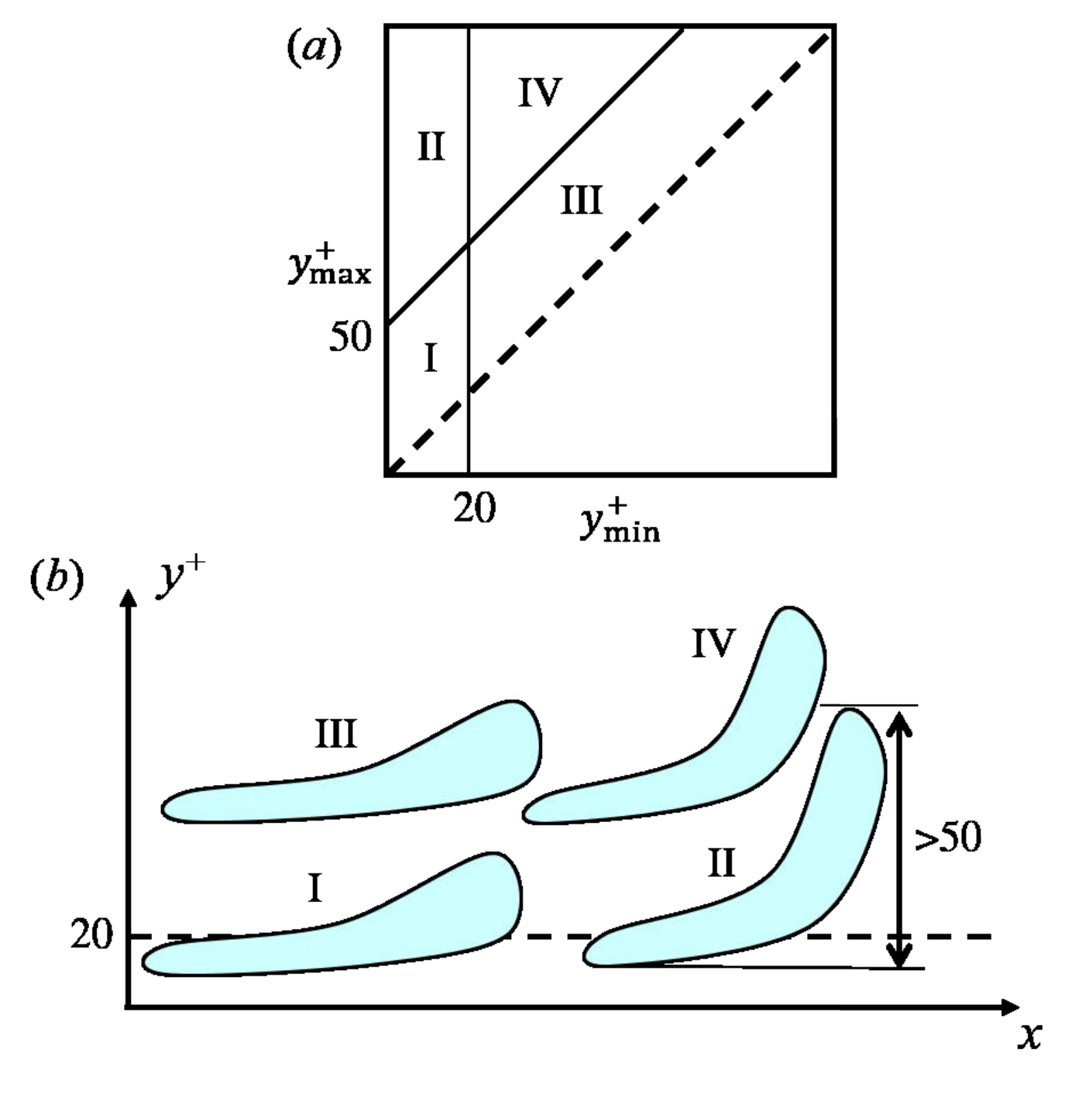}
	\caption{Schematics of vortex categorization by wall position and lift-up extent: (I) attached-flat, (II) attached-lifted, (III) detached-flat, and (IV) detached-lifted.}
	\label{fig:concept_vor_class}
\end{figure}

\citet{townsend1980structure} introduced the concepts of ``attached'' and ``detached'' vortices. Attached vortices interact closely with the wall and were believed to be responsible for the generation and transport of Reynolds stress and TKE. Detached vortices are found away from the wall and they were conjectured to be associated with the dissipation of turbulent activities~\citep{perry1995wall}.
\citet{lozano2012three} classified coherent structures into attached and detached groups based on their wall positions: structures with their bottom sticking close to the wall (i.e. $y_\text{min}^+\leq20$) were considered to be attached and the others detached.
Distinction was further made based on the wall-normal span of the structures by the same authors. In particular, ``tall-attached'' structures that extend across the channel were believed to be of particular importance in the transport of Reynolds stress.
Following the same spirit, we categorize vortices into four types based on these two metrics of vortex wall position and wall-normal span, which are both quantitatively measurable from vortex axis-lines extracted with VATIP.
Each type maps to a region in the $y_\text{max}^+\text{-}y_\text{min}^+$ coordinates (same as \cref{fig:vor_height_PDF}) as illustrated in \cref{fig:concept_vor_class}.
Type I or ``attached-flat'' vortices are those with $y_\text{min}^+\leq20$ and $l_y^+ \equiv y_\text{max}^+ - y_\text{min}^+\leq 50$. Note that the $y_\text{min}^+$ criterion measures the proximity to the wall and the $l_y^+$ criterion measures the extent of vortex lift-up. This type thus includes vortices lying flat in regions very close to the wall without strong lift-up.
These vortices are the dominant structures in the buffer layer
and are most frequently spotted in all cases (\cref{fig:vor_height_PDF}).
Type II or ``attached-lifted'' vortices satisfy $y_\text{min}^+\leq20$ and $l_y^+>50$. These vortices are generated by wall interaction but their strong lift-up allows them to efficiently transport turbulent activities between the buffer and log-law layers.
Type III or ``detached-flat'' ($y_\text{min}^+>20$ and $l_y^+\leq50$) and type IV or ``detached-lifted'' ($y_\text{min}^+>20$ and $l_y^+>50$) are similarly differentiated by their extent of lift-up and in both cases, the vortices are detached from the wall and thus less influenced by the latter.
The cut off magnitudes of $y_\text{min}^+=20$ and $l_y^+=50$ were arbitrarily chosen based on the observed distribution patterns in \cref{fig:vor_height_PDF} and our general experience with vortices in channel flow.
We have tested that changing the cut off values within a reasonable range  ($y_\text{min}^+=20\sim 40$ and $l_y^+ =35\sim 50$) would not change the following results in any significant manner.

It is necessary to clarify here that VATIP, in its current form, is an intrinsically static approach. It captures vortex instances from a frozen image of the flow field.
Therefore, categorization results according to \cref{fig:concept_vor_class} should be interpreted through the lens of ensemble statistics -- i.e., for an arbitrarily selected vortex at a random time moment, what is the probability that it is caught in a configuration belonging to one of these four types.
The method does not provide direct information on the dynamical lineage of vortices and does not track the time evolution of vortex configuration.
The category label does not carry though different times: a vortex may as well evolve into a different type at a future moment.
For instance, a classical streamwise vortex in the buffer layer would be categorized as type I, but if it lifts up at a later time, it would become type II.

\begin{figure}
	\centering
	\includegraphics[width=.99\linewidth, trim=0mm 0mm 0mm 0mm, clip]{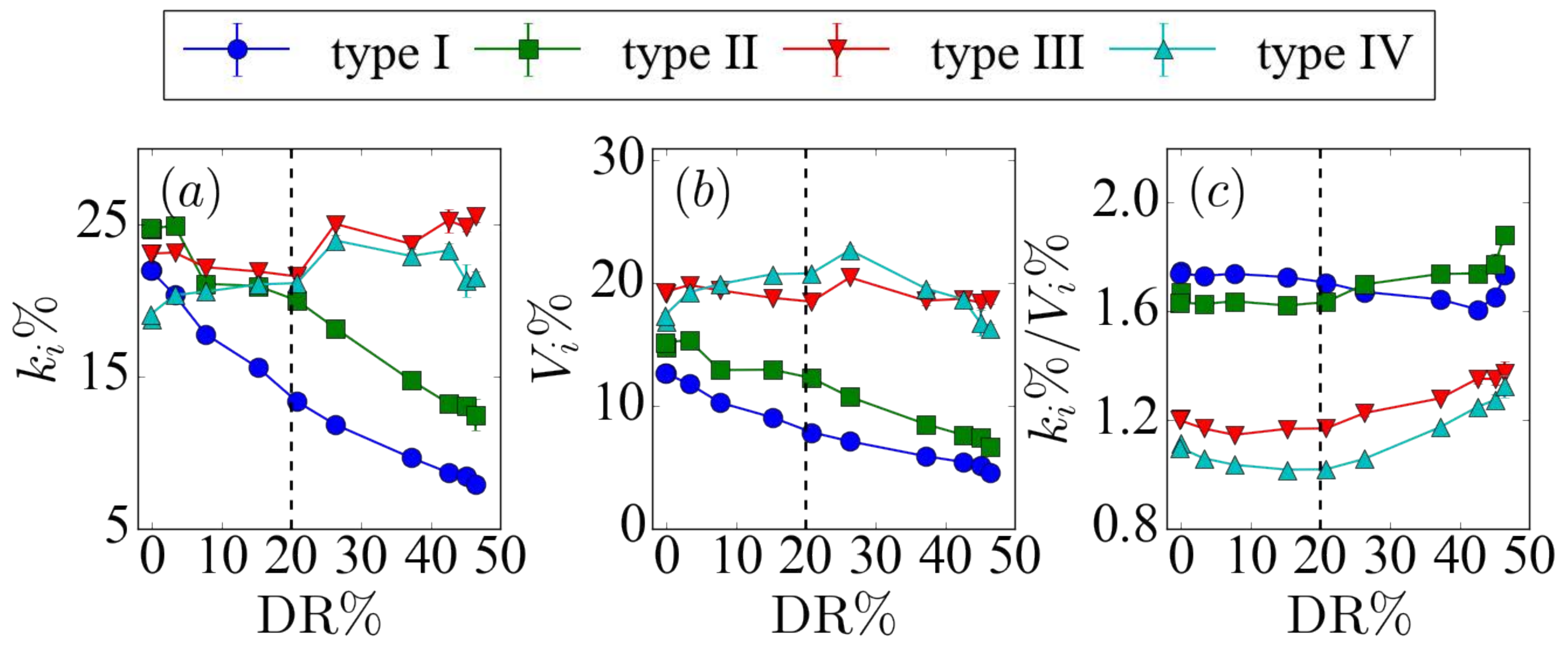}
	\caption{Distribution of TKE and volume between vortices of different types at $\mathrm{Re}_\tau=172.31$: (a) percentage of TKE contained in each type of vortices; (b) percentage of volume occupied by each type of vortex; and (c) normalized TKE density.
	Percentages are with respect to the flow domain total. Dashed line marks the LDR-HDR transition. Error bars smaller than the symbol size are not shown.}
	\label{fig:tke_DR}
\end{figure}

Polymer effects on these vortex types are quantified in \cref{fig:tke_DR} in terms of the percentage of TKE contained in all vortices of type $i$
\begin{gather}
	k_i\%\equiv \frac{k_i}{k_\text{t}}
\end{gather}
and the percentage of volume occupied by all vortices of type $i$
\begin{gather}
	V_i\%\equiv \frac{V_i}{V_\text{t}}
\end{gather}
where $k_\text{t}$ and $V_\text{t}$ are the total TKE and total volume of the flow domain, respectively. The ratio between the two
\begin{gather}
	\frac{k_i\%}{V_i\%}=\frac{k_i/V_i}{k_\text{t}/V_\text{t}}
	\label{eq:avgtkedens}
\end{gather}
is the volumetric density of TKE in vortex type $i$ normalized by the TKE density of the entire domain.
Since VATIP only renders an axis-line, instead of a three-dimensional volume, of each vortex, volumetric statistics of the vortex are calculated within a region around the axie-line.
A square with the edge length of $1.5r_\text{v}$ is drawn (in the vortex cross-sectional plane) around each axis-point (placed at the center of the square) of the vortex axis-line and regions falling into these confining squares are counted to that vortex.
In the Newtonian limit, each type takes up nearly the same share ($\approx 20\%$) of the TKE and volume. (The numbers do not add up to unity because there are regions in the flow domain not allocated to any vortex.)
With increasing $\mathrm{DR\%}$, type I (attached-flat) vortices are monotonically suppressed with dwindling shares of TKE and volume.
Type II (attached-lifted) vortices are also nearly monotonically reduced but there is a clear turning point at the LDR-HDR transition.
Reduction of type II vortices at LDR can be attributed to the general weakening of vortices (first mechanism of DR) as well as the smaller numbers of type I available as its feed. For the latter, types I and II can be viewed as different stages of the same category of attached vortices: a type I vortex may develop into a type II as it lifts up later in its lifetime~\citep{perry1995wall}. 
At HDR, lift-up suppression becomes important (\cref{fig:vor_height_PDF} and more evidences below) which leads to the faster decline of shares in type II vortices.
Polymer effects on detached (types III and IV) vortices are much subtler. There is a clear increase of TKE shares of both types at the LDR-HDR transition whereas the volume share stays roughly at the same level for all levels of DR.
As a result, the normalized TKE density (\cref{fig:tke_DR}(c)) starts to increase after the transition: i.e., as the flow reaches HDR, the relative intensity (compared with other vortex types) of detached vortices increases without them expanding in overall volume.
Since the overall turbulent intensity or the average TKE density of the flow domain $k_\text{t}/V_\text{t}$ (denominator in \cref{eq:avgtkedens}) is decreasing with $\mathrm{DR}\%$, this simply indicates that detached vortices are much less susceptible to polymer suppression, compared with attached ones, in the HDR regime.
Also, attached vortices (types I and II) are much stronger than detached ones with their TKE density more than $50\%$ higher than the latter.
At LDR, normalized density of type IV vortices are close to unity (the domain average magnitude), making them nearly not differentiable from the turbulent background.
This is consistent with observations in \cref{fig:vor_height_PDF} that this region (IV in \cref{fig:concept_vor_class}) is rarely populated by vortices.
The role of type IV is thus much less significant than the rest and it is included in our analysis for completeness only.

\begin{figure}
	\centering
	\includegraphics[width=.99\linewidth, trim=0mm 0mm 0mm 0mm, clip]{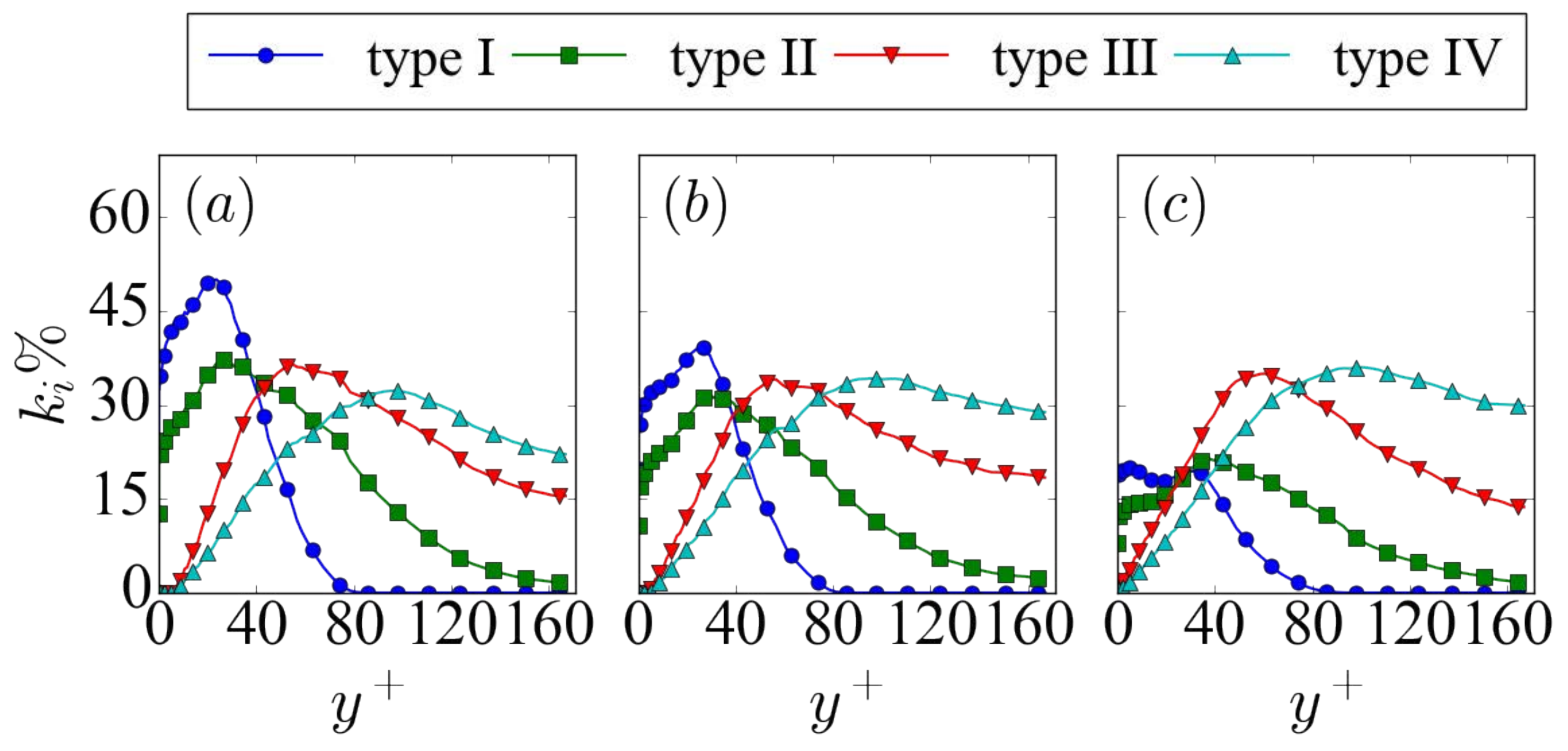}
	\caption{Distribution of turbulent kinetic energy contained in each vortex type of (a) Newtonian, (b) $\mathrm{Wi}=20$ (LDR), and (c) $\mathrm{Wi}=80$ (HDR) cases at $\mathrm{Re}_\tau=172.31$.}
	\label{fig:vor_TKE}
\end{figure}

\begin{figure}
	\centering
	\includegraphics[width=.99\linewidth, trim=0mm 0mm 0mm 0mm, clip]{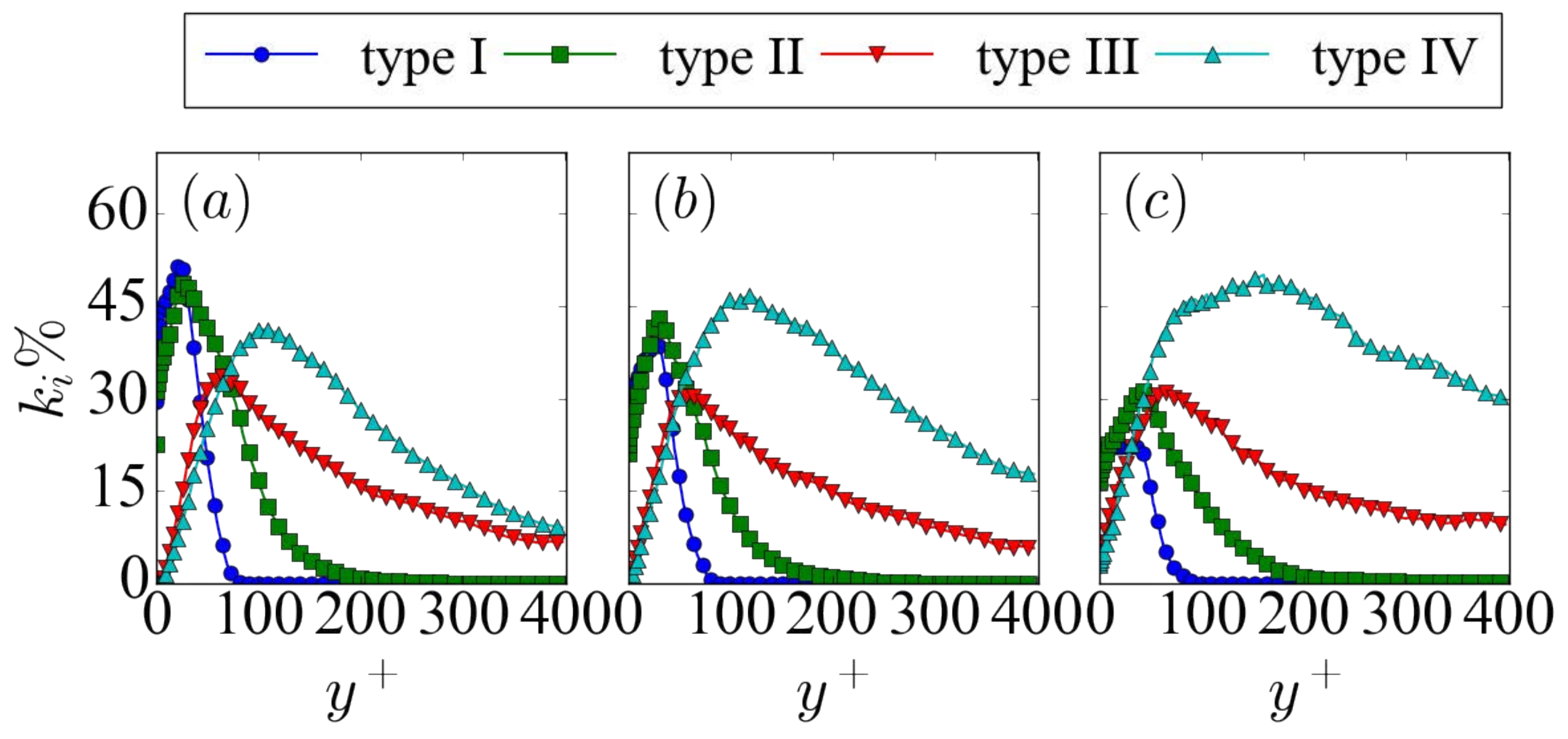}
	\caption{Distribution of turbulent kinetic energy contained in each vortex type of (a) Newtonian, (b) LDR, and (b) HDR cases at $\mathrm{Re}_\tau=400$.}
	\label{fig:vor_TKE_re80000}
\end{figure}

\Cref{fig:vor_TKE} shows the TKE share of each vortex type as a function of $y^+$ for the Newtonian, LDR ($\mathrm{Wi}=20$), and MDR ($\mathrm{Wi}=80$) cases at $\mathrm{Re}_\tau=172.31$.
Type I represents the flat-lying attached vortices and they are most predominant in the buffer layer, accounting for $50\%$ of the total TKE in the buffer layer.
Detached vortices (types III and IV) only become important in the log-law layer.
Type II, meanwhile, carries TKE across the wall layers because they originate from the wall and lift up to upper layers.
Compared with the Newtonian case, at LDR vortex type I is significantly suppressed, which corresponds to the first stage of DR effect concentrated mainly in the buffer layer. Changes in other types are much less significant.
There is a minor reduction in the type II profile within the buffer layer only, which is consistent with the earlier discussion that at LDR, type II reduction is a combined effect of general vortex weakening and reduced number of type I.
Lift-up suppression becomes important only at HDR where reduction of the type II profile in the log-law region becomes significant (as type I share continues to drop).
Meanwhile, profiles for detached vortices (types III and IV) are slightly raised.

The same observations are largely preserved at the higher $\mathrm{Re}_\tau=400$ (\cref{fig:vor_TKE_re80000}).
Compared with the lower $\mathrm{Re}$ case, increasing $\mathrm{Re}$ leads to an overall increase of lifted vortices, both attached (type II) and detached (type IV). This is consistent with the previous finding in Newtonian turbulence that lifted-up three-dimensional vortices (e.g., hairpins) become more prevalent at higher $\mathrm{Re}$~\citep{Zhu_Xi_JFM2019}.
For both $\mathrm{Re}$, attached vortices (types I and II) are contained within roughly the same wall layers in inner units: type I is found at $y^+\lesssim100$ and type II shows highest TKE at $y^+\approx 30$ and its upper end extends close to $y^+\approx 200$.
Meanwhile, detached vortices (types III and IV) are less contained and spread to the highest $y^+$ available at each $\mathrm{Re}$. The position of peak TKE, however, is still comparable in inner units at different $\mathrm{Re}$.
The effect of increasing $\mathrm{Wi}$ and comparison between different stages of DR remain the same between these two $\mathrm{Re}$.

In summary, analysis of vortices of different types shows that polymers mainly suppress attached vortices. This effect is confined to the buffer layer at LDR.
Polymer effects on TKE distribution in the log-law region becomes important only at HDR because of their suppression of vortex lift-up (evidences in \cref{fig:vor_height_PDF} and also below), which reduces the turbulent momentum transfer between wall layers and results in the changing flow profiles in the log-law layer.

\subsection{Vortex shape distribution at different stages of DR}\label{Sec_vor_type}

\begin{figure}
	\centering
	\includegraphics[width=.99\linewidth, trim=0mm 0mm 0mm 0mm, clip]{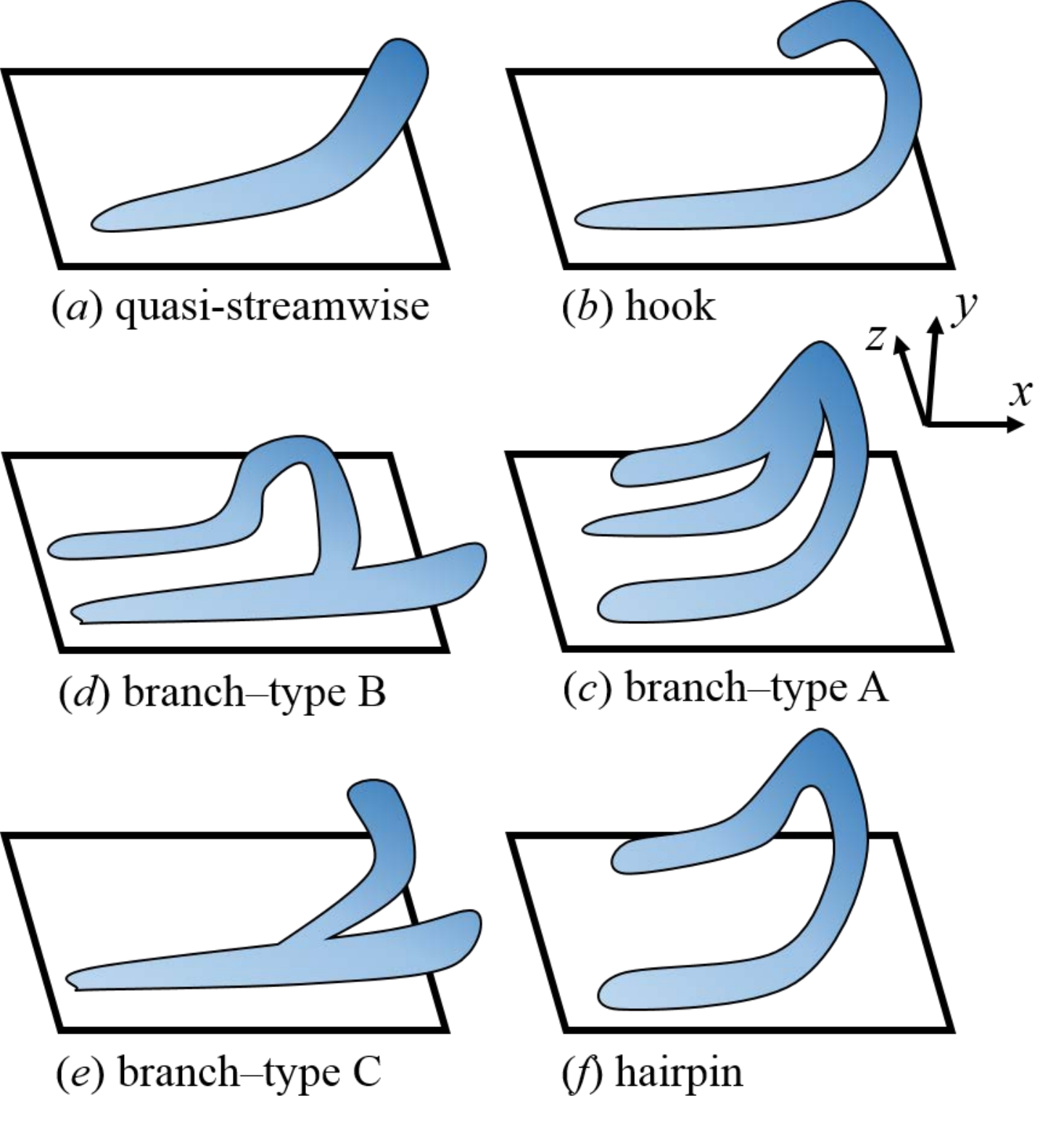}
	\caption{Schematic illustrations of major vortex types by shape.}
	\label{fig:class_flow} 
\end{figure}

\begin{figure}
	\centering
	\includegraphics[width=0.8\linewidth, trim=0mm 0mm 0mm 0mm, clip]{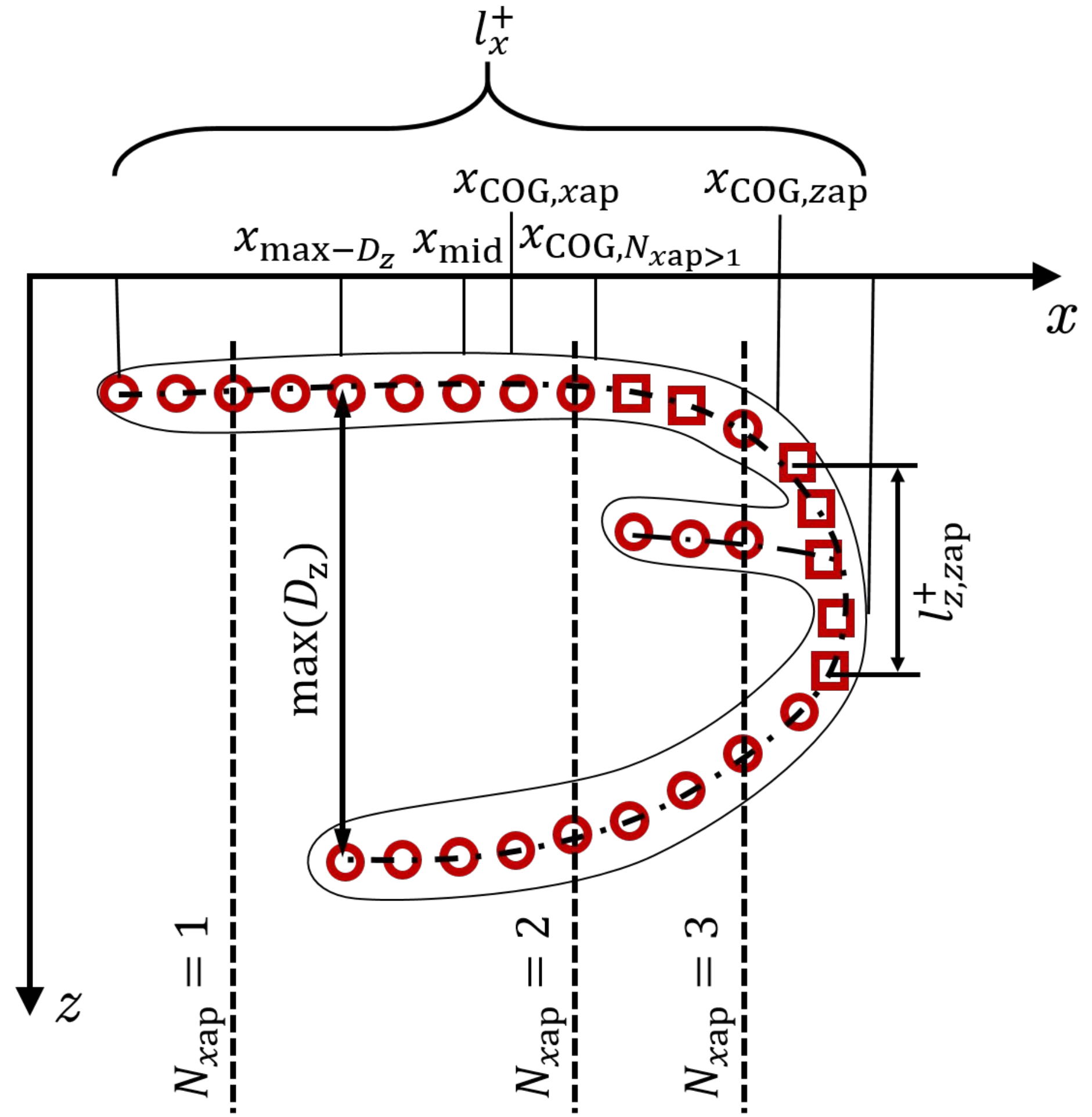}
	\caption{Definitions of vortex metrics used in the classification of their shapes ($xz$-plane projection). Circular and square markers represent $x$- and $z$-axis-points, respectively.}
	\label{fig:class_para}
\end{figure}

\begin{table}
	\begin{center}
		\begin{tabular}{p{28ex}|cccccccc}
			\multirow{2}{*}{Condition} &
			\multirow{2}{*}{Frag.}  &
			\multirow{2}{*}{Stream.}  &
			\multirow{2}{*}{Hook} &
			\multicolumn{3}{c}{Branch} &
			\multirow{2}{*}{Hairpin}\\
			&&&&A  &B  &C  &\\
		\hline
			$l_x^+<50$	&  T  &  F	&  F
				&  F  &  F	&  F  &  F\\
			$\max(l^+_{z,z\text{ap}}) < 25$  &  -  &  T  &  F
				&  F  &  F  &  F  &  F\\
			$P_x(N_{x\text{ap}}=1)>80\%$  &  -  &  -  &  T
				&  F  &  F  &  F  &  F\\
			$\frac{P_x(N_{x\text{ap}}>2)}{P_x(N_{x\text{ap}}>1)}>50\%$  &  -  &  -
				&  -  &  T  &  F  &  F  &  F\\
			\begin{tabular}{l}
			$x_{\text{COG},x\text{ap}}<x_\text{mid}$ or\\ $x_{\text{COG},z\text{ap}}<x_\text{mid}$
			\end{tabular}	&  -  &  -  &  - 
				&  -  &  T  &  F  &  F\\
			$x_{\text{max-}D_z}>1.5x_{\text{COG},N_{x\text{ap}}>1}$  &  -  &  -  &  -
				&  -  &  -  &  T  &  F\\
		\end{tabular}
		\caption{Vortex classification criteria based on geometric metrics of the axis-line.}
		\label{tab:vor_class}
	\end{center}	
\end{table}

Analysis so far has been focused on the size, wall position, and lift-up status of vortices without considering their topological shape.
Determination of the latter by a computer code requires a set of quantitative criteria on the vortex geometry.
We will adopt the vortex classification procedure proposed in \citet{Zhu_Xi_JFM2019} based on measurements of axis-lines extracted by VATIP.
Like before, we will only recapitulate the approach at the conceptual level here and refer the readers to \citet{Zhu_Xi_JFM2019} for implementation details.
Vortices are classified into six major types illustrated in \cref{fig:class_flow} based on quantitative metrics defined in \cref{fig:class_para}.
Criteria for differentiating different types are summarized in \cref{tab:vor_class}.

The classification is done by a series of binary decisions. First, it differentiates fragments from substantial vortices by requiring the streamwise length $l^+_x$ to be at least 50 for the latter.
Second, it identifies quasi-streamwise vortices by measuring the length of the longest spanwise segment in the axis-line $\max(l^+_{z,z\text{ap}})$. (Spanwise segments are those consisting of a string of connected $z$-axis-points.) Those with $\max(l^+_{z,z\text{ap}})<25$ are considered to not have a substantial spanwise arm to be considered a hairpin or any other branched type. Note that streamwise vortices that become highly lifted up are still considered in this class because there is no restriction on wall-normal segments.
Third, the hook type, which can be viewed as an incomplete hairpin with only one fully developed leg, is identified by counting the number of $x$-axis-points in each $yz$-planes $N_{x\text{ap}}$. If more than 80\% of the $yz$-planes spanned by the axis-line has only 1 $x$-axis-point, it is determined that the vortex is dominated by one streamwise leg. (In \cref{tab:vor_class}, $P_x(N_{x\text{ap}}=1)$ represents the percentage of $yz$-planes that satisfy the condition of $N_{x\text{ap}}=1$).
Fourth, the remaining unsorted group are either hairpins or irregularly branched vortices with some features of hairpins but do not conform to their canonical $\Omega-$shape.
A commonly-seen type is a hairpin-like structure with 3 or more legs. These vortices can be formed when a hairpin is merged with another vortex in highly crowded vortex packets.
The third leg is considered to be substantial if the number of $yz$-planes containing more than 2 $x$-axis-points (intersected by 3 or more legs) is more than that of those with only $2$ (planes intersected by two legs). These vortices are classified as branch type A.
Fifth, the branch type B (\cref{fig:class_flow}) can be formed when a side arm of the streamwise vortex lifts up and is dragged sideways by the spanwise flow to form an arc and, sometimes, another leg. It is similar to a hairpin except that the head or arc of the vortex is not found near the downstream end but somewhere in the middle. The vortex head is considered to be significantly away from the downstream end if the $x$-coordinate of the center of gravity (COG) of either all $x$-axis-points $x_{\text{COG},x\text{ap}}$ or all $z$-axis-points $x_{\text{COG},x\text{ap}}$ is upstream of the middle point of the entire $x$-span ($x_\text{mid}\equiv(x_\text{max}+x_\text{min})/2$).
Sixth, branch type C is formed in a similar manner except that the side arm is stretched by the streamwise flow first before lifting up, creating a branch that opens towards the downstream direction. In this case, the $x$-coordinate with the maximum spanwise span $D_z$ is found near the downstream end.
The quantitative criterion is to compare this coordinate $x_{\text{max-}D_z}$ with that of the COG. of the branched portion (i.e., those where $N_{x\text{ap}}>1$) $x_{\text{COG},N_{x\text{ap}}>1}$ multiplied by 1.5.
Finally, after removing all irregularly branched configurations, the rest are considered to be sufficiently close to the canonical $\Omega$-shape and classified as hairpins.

In summary, after removing the fragments, quasi-streamwise vortices, and hooks from the pool, the algorithm identifies hairpins by removing all other branched types with significant deviation from the canonical $\Omega$-shape.
There is obviously some arbitrariness in how the branch types (A, B, and C) are defined and how the cutoffs are chosen (i.e., when is a deviation big enough to disqualify a vortex as a hairpin).
Fortunately, at least for this study, this is nothing more than a taxonomic issue. For practical purposes, none of the trends we will discuss below show any difference between hairpins and other branches (types A, B, and C).
This is not surprising: within our current limited knowledge of vortex dynamics, all these branches seem to be formed in a similar manner as hairpins. Their existence is merely an inevitability of the irregular nature of turbulent dynamics.
For this reason, we will use one umbrella term ``hairpin-like'' vortices for all these branched structures (including canonical hairpins).

\begin{figure}
	\centering
	\includegraphics[width=.95\linewidth, trim=0mm 24mm 0mm 0mm, clip]{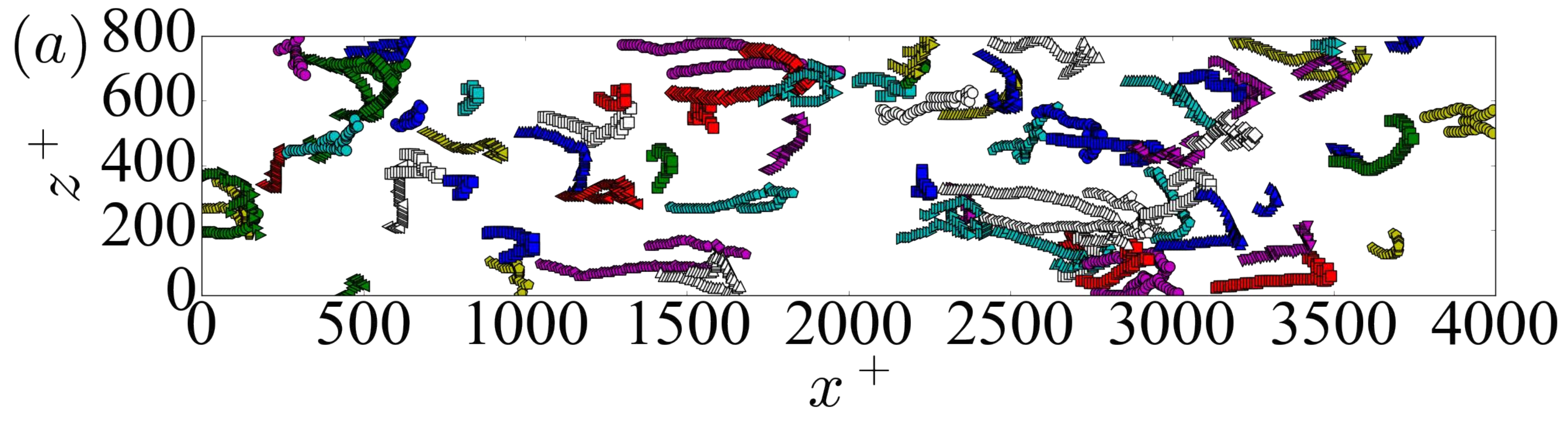}
	\includegraphics[width=.95\linewidth, trim=0mm 24mm 0mm 0mm, clip]{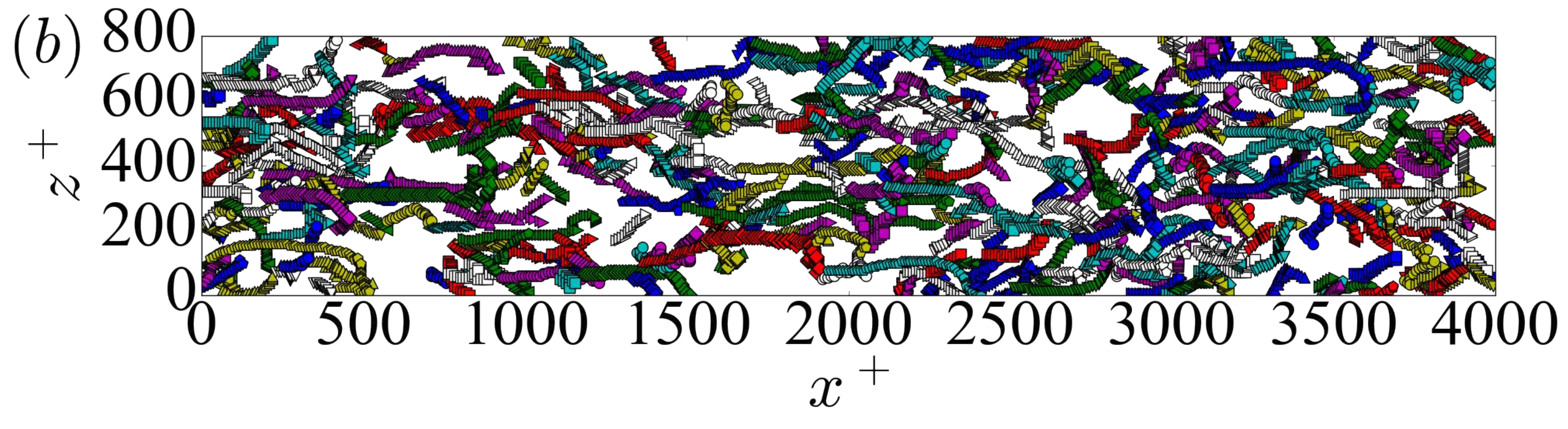}
	\includegraphics[width=.95\linewidth, trim=0mm 24mm 0mm 0mm, clip]{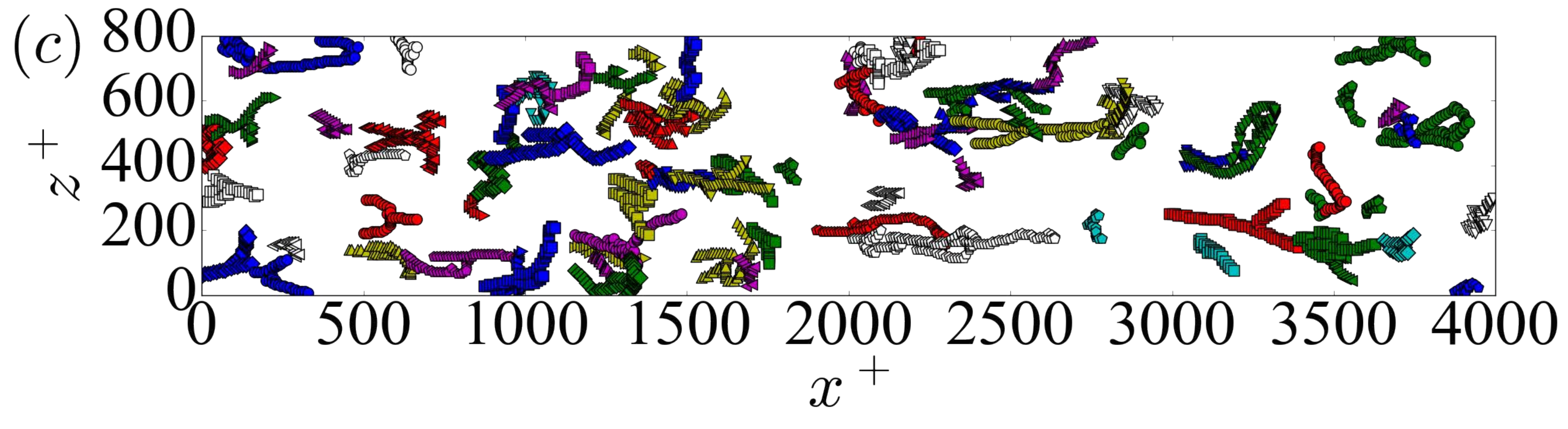}
	\includegraphics[width=.95\linewidth, trim=0mm 24mm 0mm 0mm, clip]{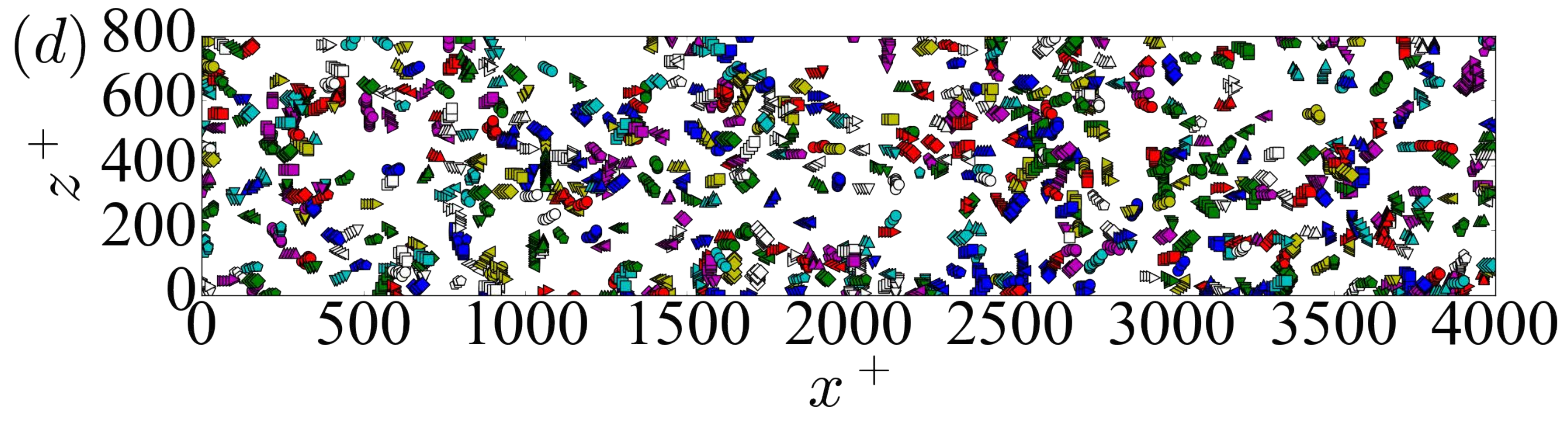}
	\includegraphics[width=.95\linewidth, trim=0mm 0mm 0mm 0mm,  clip]{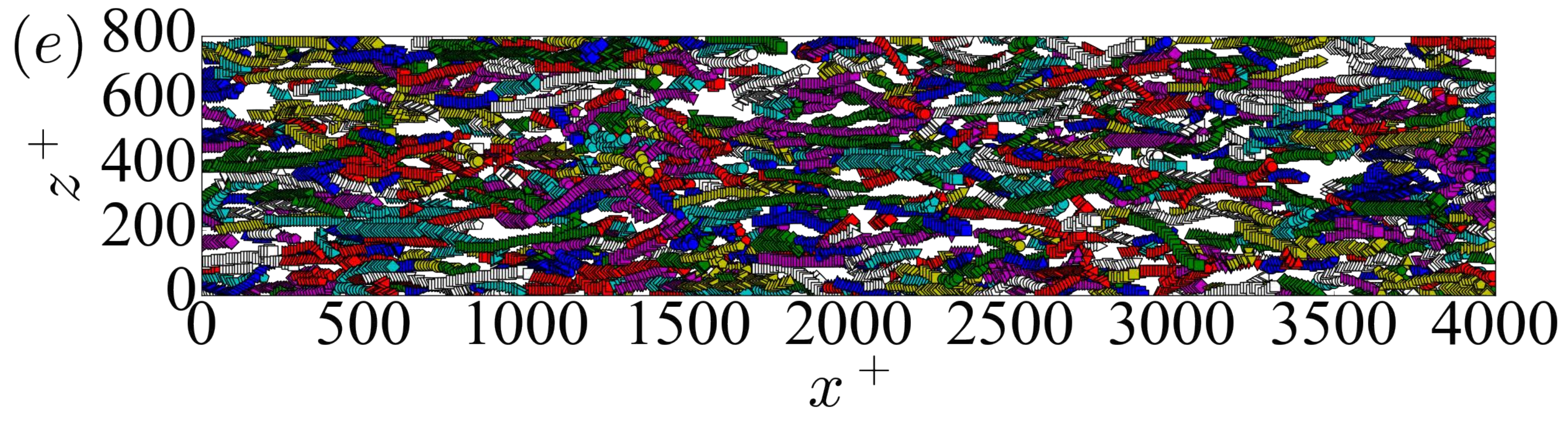}
	\caption{Axis-lines of vortices of different shapes extracted by VATIP in a typical snapshot at $\mathrm{Re}_\tau=172.31$ and $\mathrm{Wi}=20$ (LDR): (\textit{a}) hairpin, (\textit{b}) hook, (\textit{c}) branch, (\textit{d})fragment and (\textit{e}) quasi-streamwise vortices. Different vortices are represented by different colors and markers.
	Viewed from above the channel and the projection includes vortices at all $y$ positions.}
	\label{fig:vor_class_visu1}
\end{figure}

\begin{figure}
	\centering
	\includegraphics[width=.95\linewidth, trim=0mm 24mm 0mm 0mm, clip]{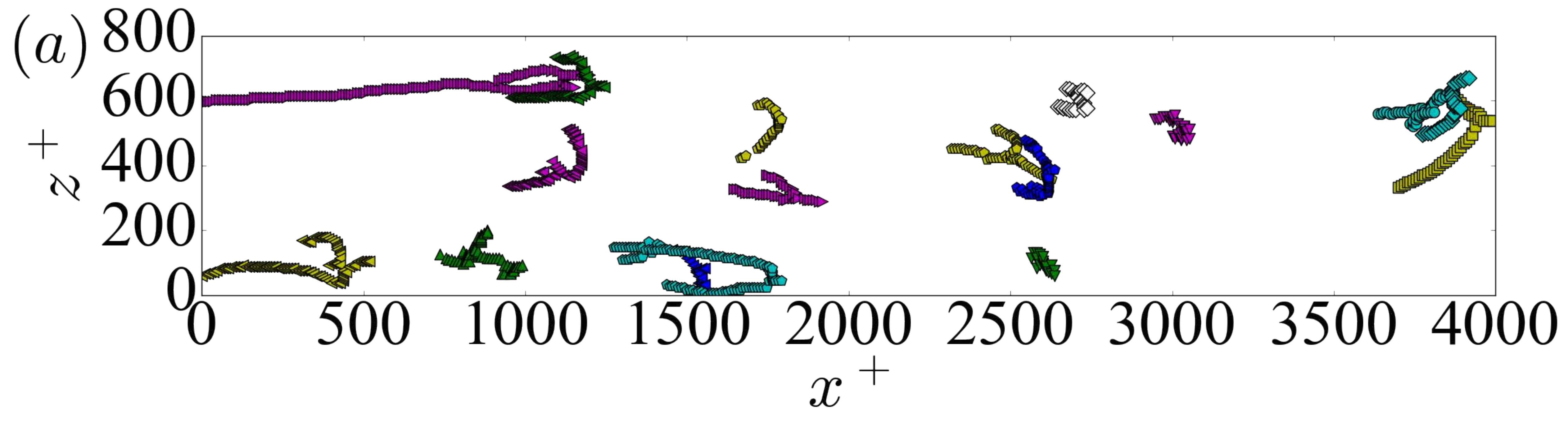}
	\includegraphics[width=.95\linewidth, trim=0mm 24mm 0mm 0mm, clip]{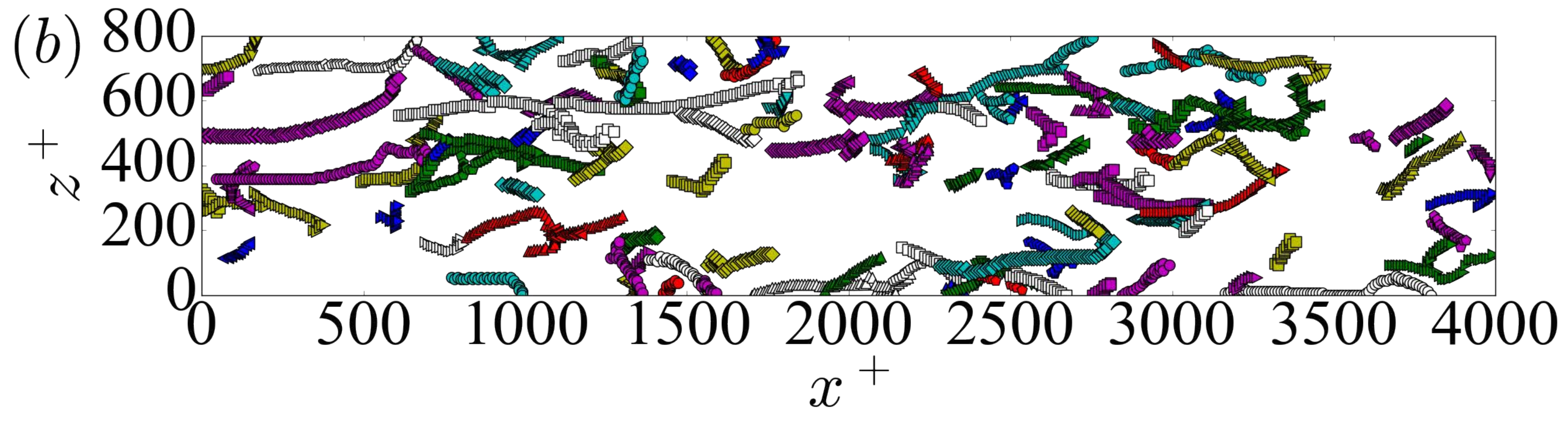}
	\includegraphics[width=.95\linewidth, trim=0mm 24mm 0mm 0mm, clip]{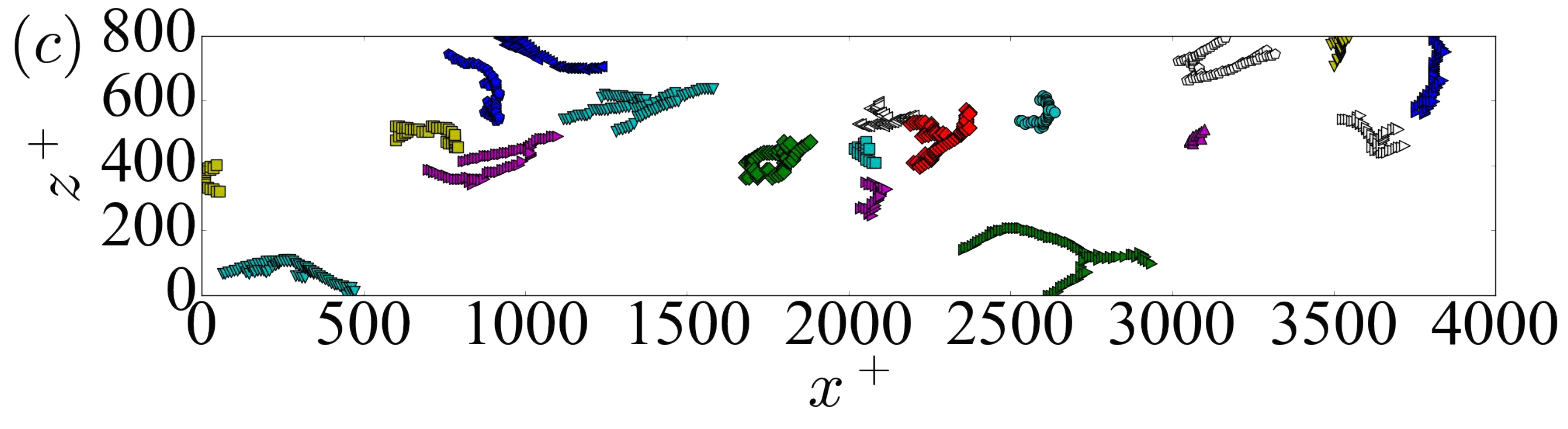}
	\includegraphics[width=.95\linewidth, trim=0mm 24mm 0mm 0mm, clip]{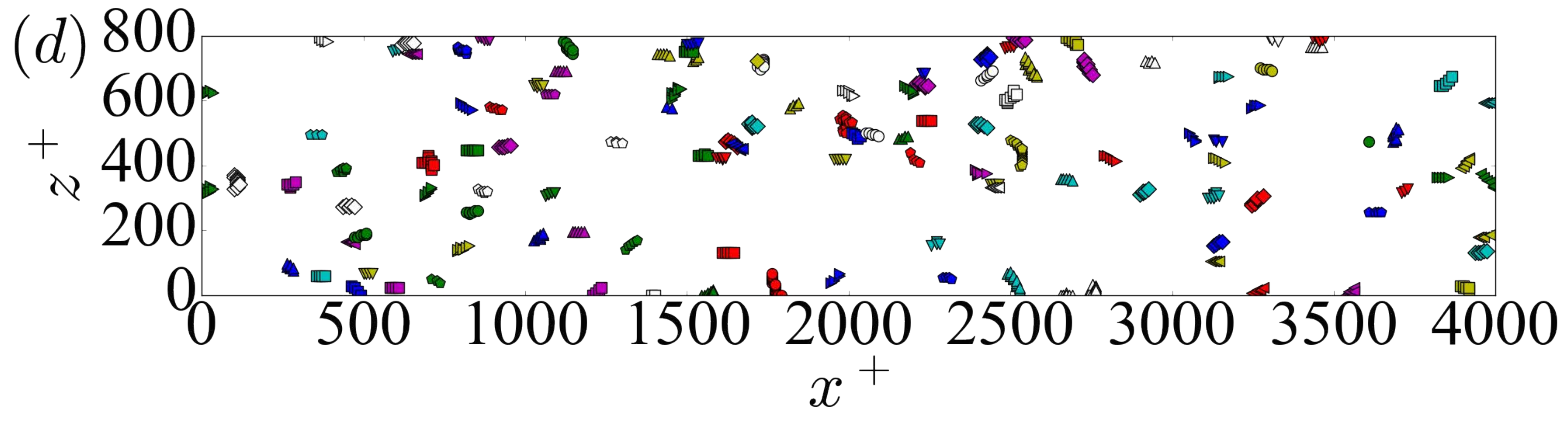}
	\includegraphics[width=.95\linewidth, trim=0mm 0mm 0mm 0mm,  clip]{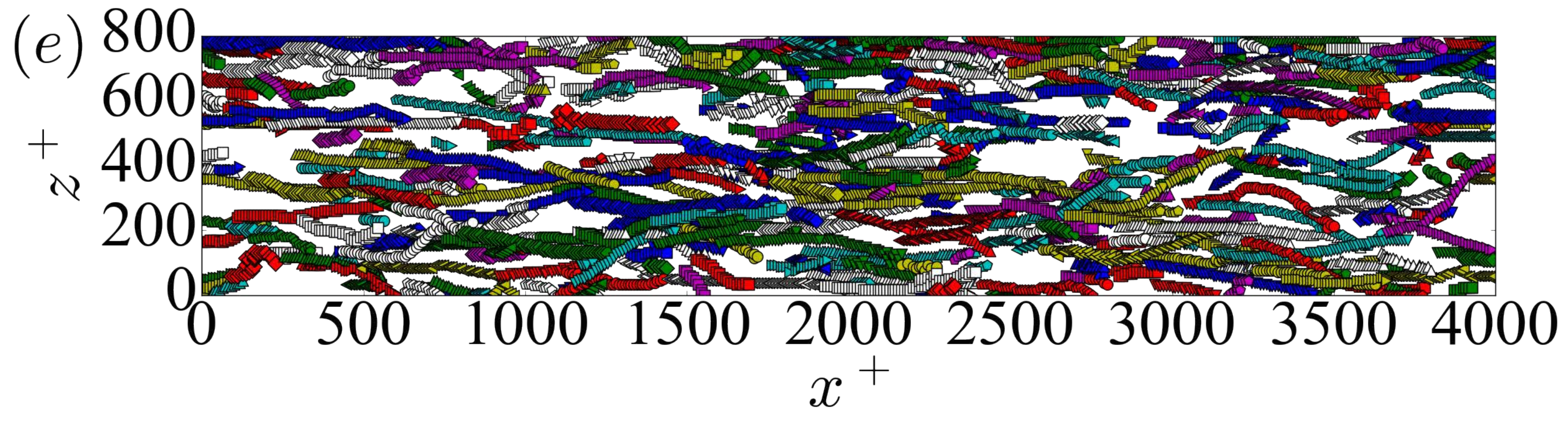}
	\caption{Axis-lines of vortices of different shapes extracted by VATIP in a typical snapshot at $\mathrm{Re}_\tau=172.31$ and $\mathrm{Wi}=80$ (HDR): (\textit{a}) hairpin, (\textit{b}) hook, (\textit{c}) branch, (\textit{d})fragment and (\textit{e}) quasi-streamwise vortices. Different vortices are represented by different colors and markers.
	Viewed from above the channel and the projection includes vortices at all $y$ positions.}
	\label{fig:vor_class_visu2}
\end{figure}

Vortex axis-lines of all these types, at $\mathrm{Re}_\tau=172.31$, are shown in \cref{fig:vor_class_visu1,fig:vor_class_visu2} for one typical snapshot at LDR ($\mathrm{Wi}=20$) and HDR ($\mathrm{Wi}=80$) each. (The ``branch'' case includes all three types, A, B, and, C and we make no further attempt to differentiate these groups.)
In both cases, near-wall quasi-streamwise vortices are the most prevalent type of vortex structure in the flow field.
However, in the LDR case, a considerable number of curved vortices are observed, including many well-defined hairpins (\cref{fig:vor_class_visu1}(a)) and other branches (\cref{fig:vor_class_visu1}(c)). They are however significantly outnumbered by the strongly asymmetric hooks (i.e., one-legged hairpins).
Observation in Newtonian flow is similar~\citep{Zhu_Xi_JFM2019}. Indeed, it has been long believed that complete well-defined hairpins are not the most likely configuration and incomplete and asymmetric hairpins (hooks) are the norm~\citep{robinson1991coherent}.
(A ``forest'' of nearly symmetric hairpins were observed in the DNS by \citet{wu2009direct} in boundary layer flow, which is different than the channel flow here.)
At HDR, all three-dimensional curved vortices (hairpins, branches, and hooks) are significantly reduced. This again is explained by the suppression of vortex lift-up which is required for their formation.
In addition, fragments also become drastically reduced in the HDR case.
\RevisedText{Since fragments are often generated in the aftermath of bursting and can be viewed as the debris of broken vortices, their reduction at HDR is also consistent with the hypothesis in \citet{Zhu_Xi_JNNFM2018}: i.e., the suppression of vortex lift up prevents its further bursting and the subsequent generation of small-scale turbulent fluctuations (which can trigger instabilities elsewhere in the domain), leaving turbulence at HDR to be dominated by a different vortex regeneration mechanism.}

\begin{figure}
	\centering
	\includegraphics[width=0.98\linewidth, trim=0mm 0mm 0mm 0mm, clip]{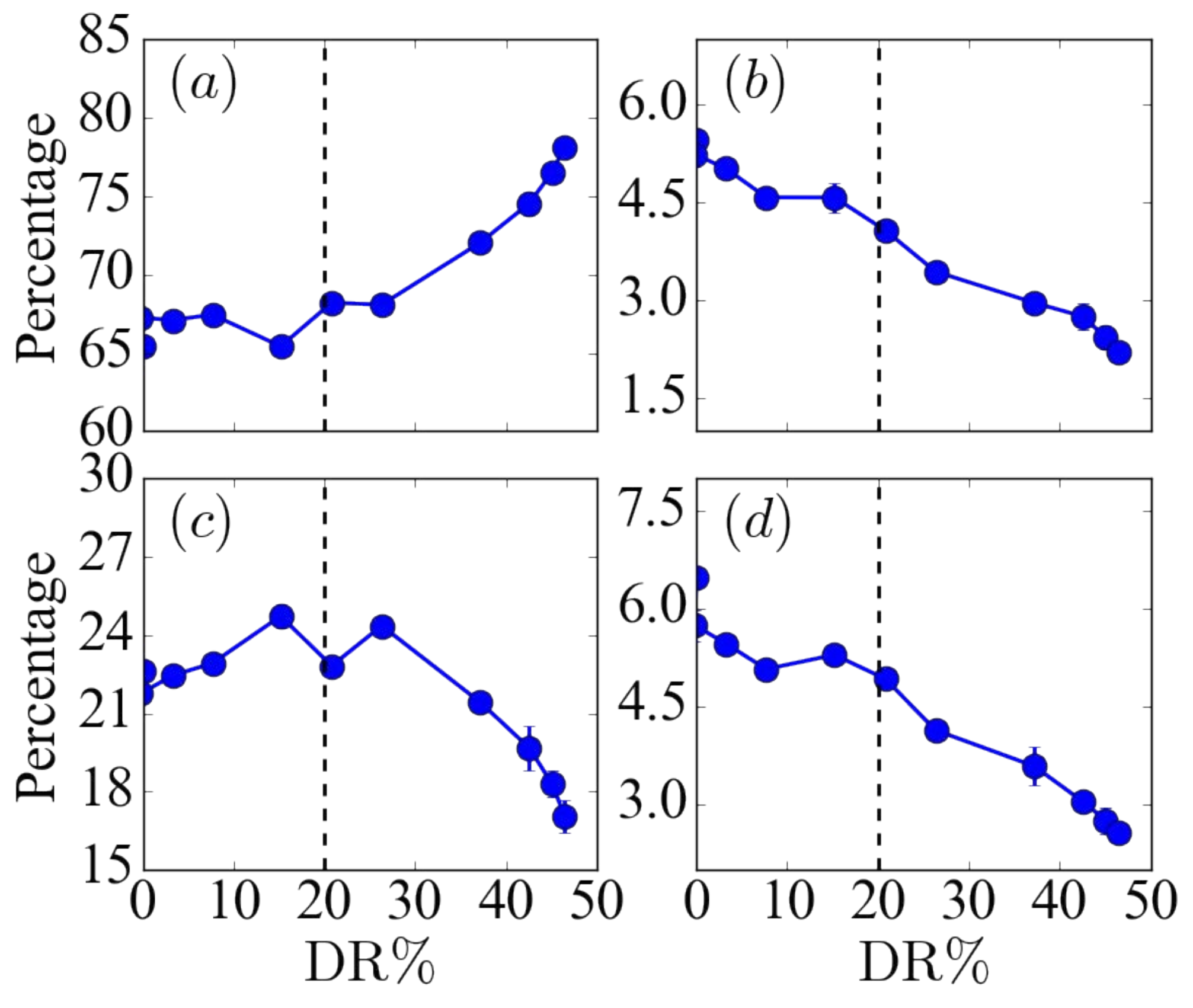}
	\caption{Number percentage of vortices of different shapes at $\mathrm{Re}_\tau=172.31$: (a) quasi-streamwise, (b) hairpin, (c) hook and (d) branch vortices.
	Dashed line marks the LDR-HDR transition.
	Error bars smaller than the symbol size are not shown.}
	\label{fig:vor_pec}
\end{figure}

Percentages of vortices of these shape types at $\mathrm{Re}_\tau=172.31$ are plotted in \cref{fig:vor_pec} as functions of $\mathrm{DR\%}$.
Changes during LDR are relatively small. The fraction of streamwise vortices remains nearly invariant.
Some subtle changes are observed in curved vortices in a small region after the onset ($\mathrm{DR}\%<5\%$): the shares taken by hairpin-like vortices (panels (b) and (d)) drop slightly, which is compensated by an increase in hooks (panel (c)).
This again shows that during this first stage of DR, polymers have an across-the-board vortex weakening effect. It suppresses all types of vortices~\citep{Zhu_Xi_JNNFM2018, DeAngelis_Piva_CompFl2002, dubief2005new, kim2007effects} without tipping the balance between them.
Changes between hooks and hairpin-like vortices can be well explained considering that some of the latter type are turned into hooks as their legs are shortened and trimmed by the polymer stress, but they remain distinguishable from quasi-streamwise ones with their spanwise arc and strong lift-up angle.
Once HDR starts, all these highly curved vortices (hairpins, branches, hooks) decline sharply as the quasi-streamwise type makes inroads into their shares.
This again can be explained by the suppression of vortex lift-up that generates these curved three-dimensional vortices during this second stage of DR. Without lift-up, streamwise vortices is stabilized near the wall and becomes elongated over time as seen in \cref{fig:vor_len,fig:vor_class_visu2}.

\begin{figure}
	\centering
	\includegraphics[width=.99\linewidth, trim=0mm 0mm 0mm 0mm, clip]{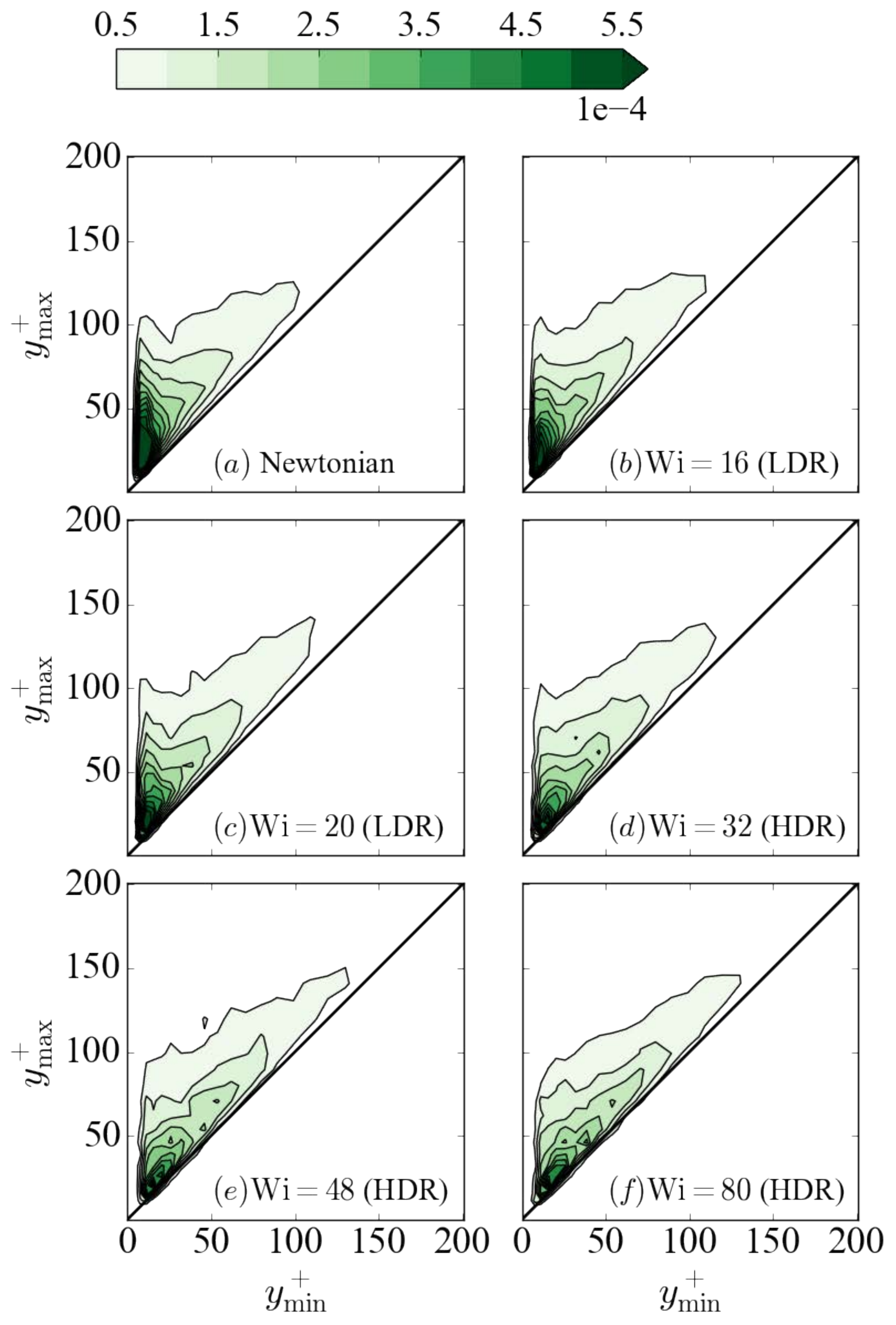}
	\caption{Joint PDFs of the wall-normal positions of the head and tail/legs of quasi-streamwise vortices, as measured respectively by the maximum and minimum $y^+$ coordinates of each vortex axis-line, at different $\mathrm{Wi}$ ($\mathrm{Re}_\tau=172.31$).}
	\label{fig:vor_height_PDF_str}
\end{figure}

\begin{figure}
	\centering
	\includegraphics[width=.99\linewidth, trim=0mm 0mm 0mm 0mm, clip]{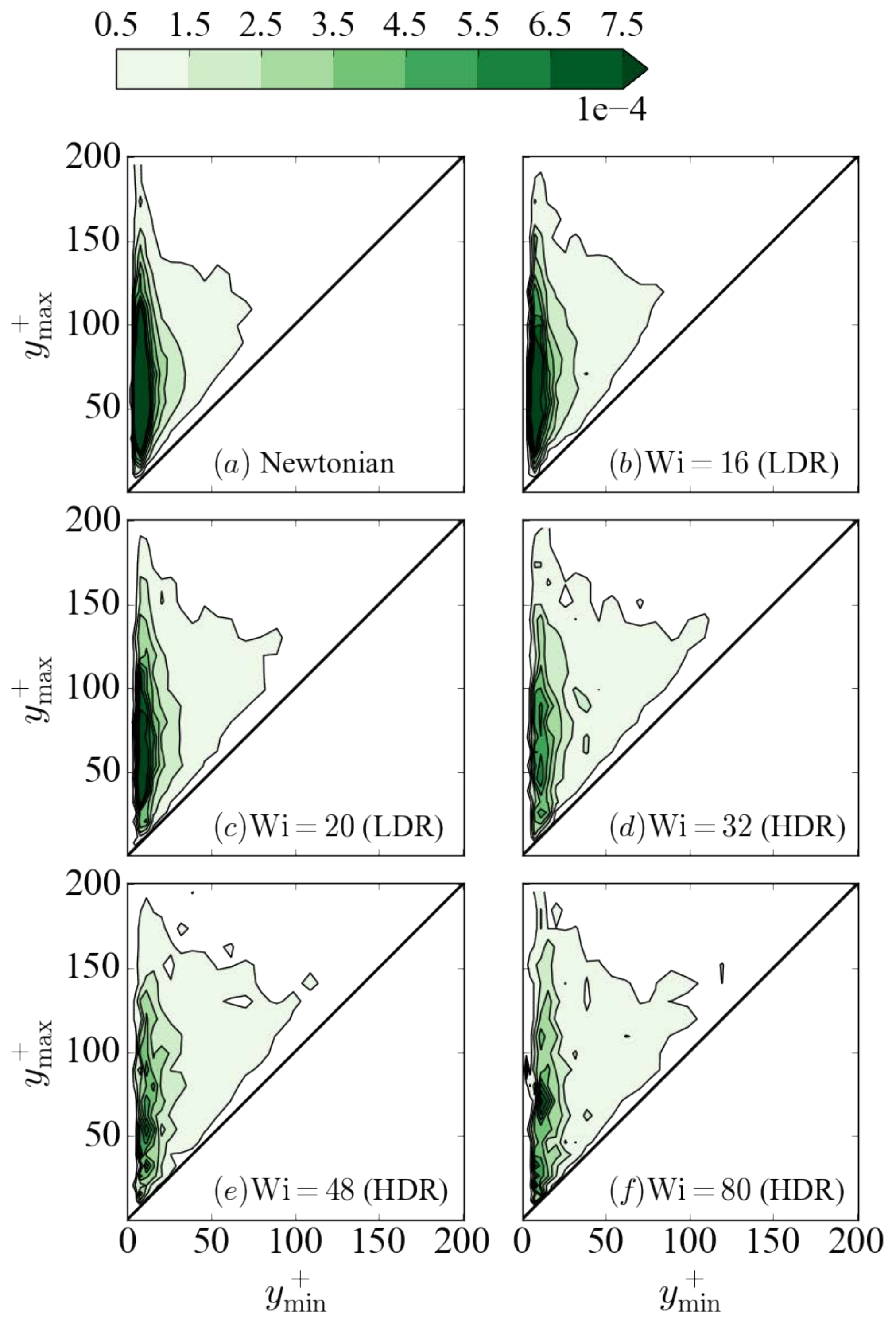}
	\caption{Joint PDFs of the wall-normal positions of the head and tail/legs of hairpin-like vortices (hairpins and branches), as measured respectively by the maximum and minimum $y^+$ coordinates of each vortex axis-line, at different $\mathrm{Wi}$ ($\mathrm{Re}_\tau=172.31$).}
	\label{fig:vor_height_PDF_cur}
\end{figure}

We now revisit the vortex position and lift-up status analysis (see \cref{fig:vor_height_PDF}) but consider vortices of different shapes in separate categories.
\Cref{fig:vor_height_PDF_str,fig:vor_height_PDF_cur} show the joint PDFs of vortex head and tail/legs positions for quasi-streamwise and hairpin-like vortices for the lower $\mathrm{Re}_\tau=172.31$.
Distribution patterns are drastically different between these two categories.
For Newtonian flow, quasi-streamwise vortices are mostly found in the lower-left corner and belong to the attached-flat class or type I (\cref{fig:concept_vor_class}).
Some of them lift up and form a thin band near the ordinate: i.e. type II attached-lifted vortices.
A diagonal band is also noticeable which corresponds to type III detached-flat vortices.
By contrast, hairpin-like vortices (including branches) are predominantly type II (attached-lifted) which originate from the wall (legs) but lift high up into the upper layers (the head or arc of the hairpin).
At LDR, the contours remain similar to the Newtonian limit for both quasi-streamwise and hairpin-like vortices.
Earlier observation of the decline of TKE shares contained in types I and II (\cref{fig:tke_DR,fig:vor_TKE,fig:vor_TKE_re80000}) are thus results of the weakening of these vortices rather than any fundamental change in their distribution pattern.
This starts to change at HDR. For quasi-streamwise vortices (\cref{fig:vor_height_PDF_str}), the slim vertical distribution band (type II) disappears as HDR starts, which is accompanied by a distinct shift of the concentration center towards the diagonal. This is a clear indication that polymers start to suppress the lift-up of these vortices and stabilize them in the streamwise direction.
Expansion of streamwise vortex distribution to higher $y^+$ (more detached) positions along the diagonal is comprehensible considering that drag-reducing polymers are known to enlarge the diameter of vortex tubes~\citep{sureshkumar1997direct, DeAngelis_Piva_PRE2003, li2006influence, Xi_Graham_JFM2010}, which inevitably raises the positions of their axis-lines. 
By contrast, hairpin-like vortices stay mainly in the type II region for the whole range of DR (\cref{fig:vor_height_PDF_cur}). 

\begin{figure}
	\centering
	\includegraphics[width=.99\linewidth, trim=0mm 0mm 0mm 0mm, clip]{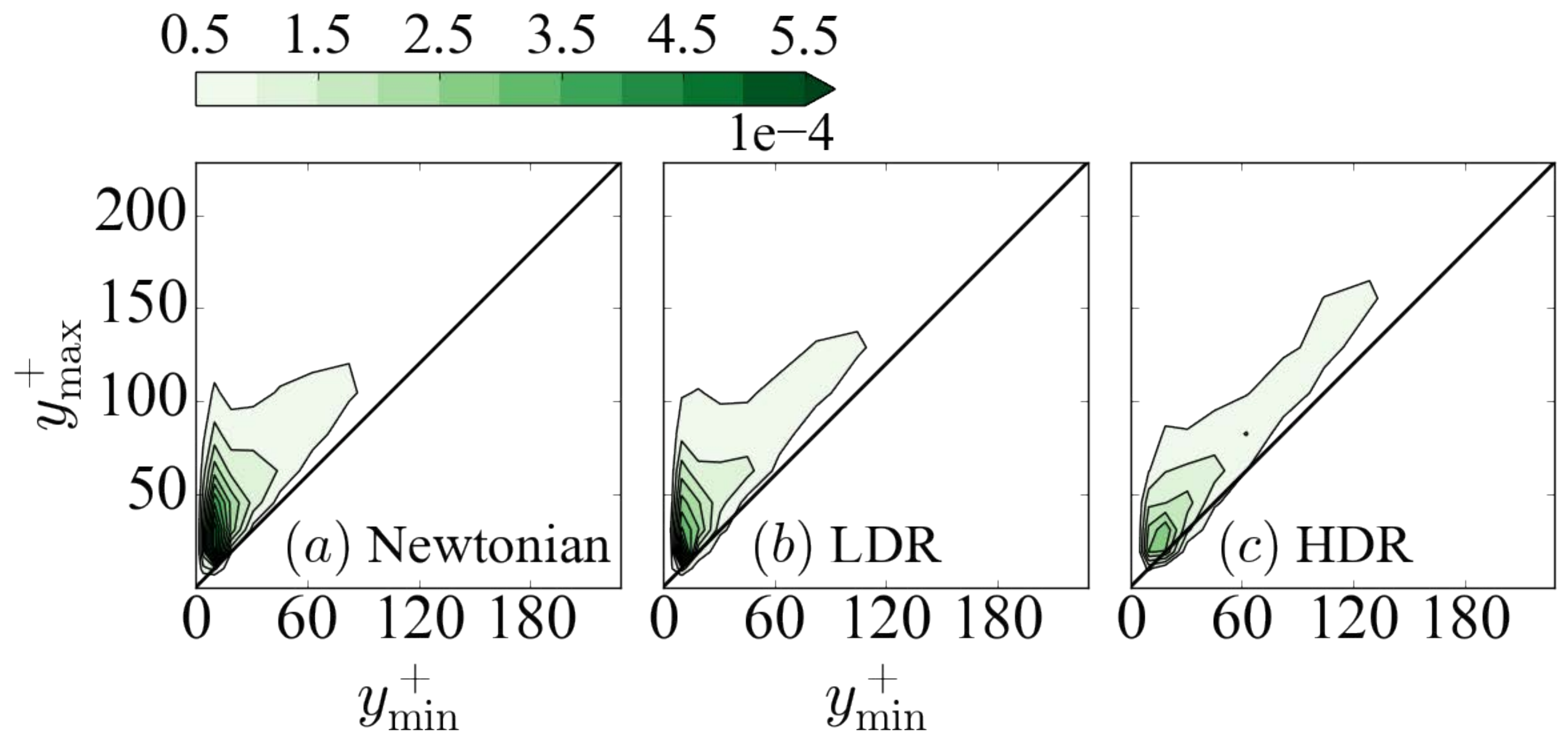}
	\caption{Joint PDFs of the wall-normal positions of the head and tail/legs of quasi-streamwise vortices, as measured respectively by the maximum and minimum $y^+$ coordinates of each vortex axis-line, at different $\mathrm{Wi}$ ($\mathrm{Re}_\tau=400$).}
	\label{fig:vor_height_PDF_str_re80000}
\end{figure}

\begin{figure}
	\centering
	\includegraphics[width=.99\linewidth, trim=0mm 0mm 0mm 0mm, clip]{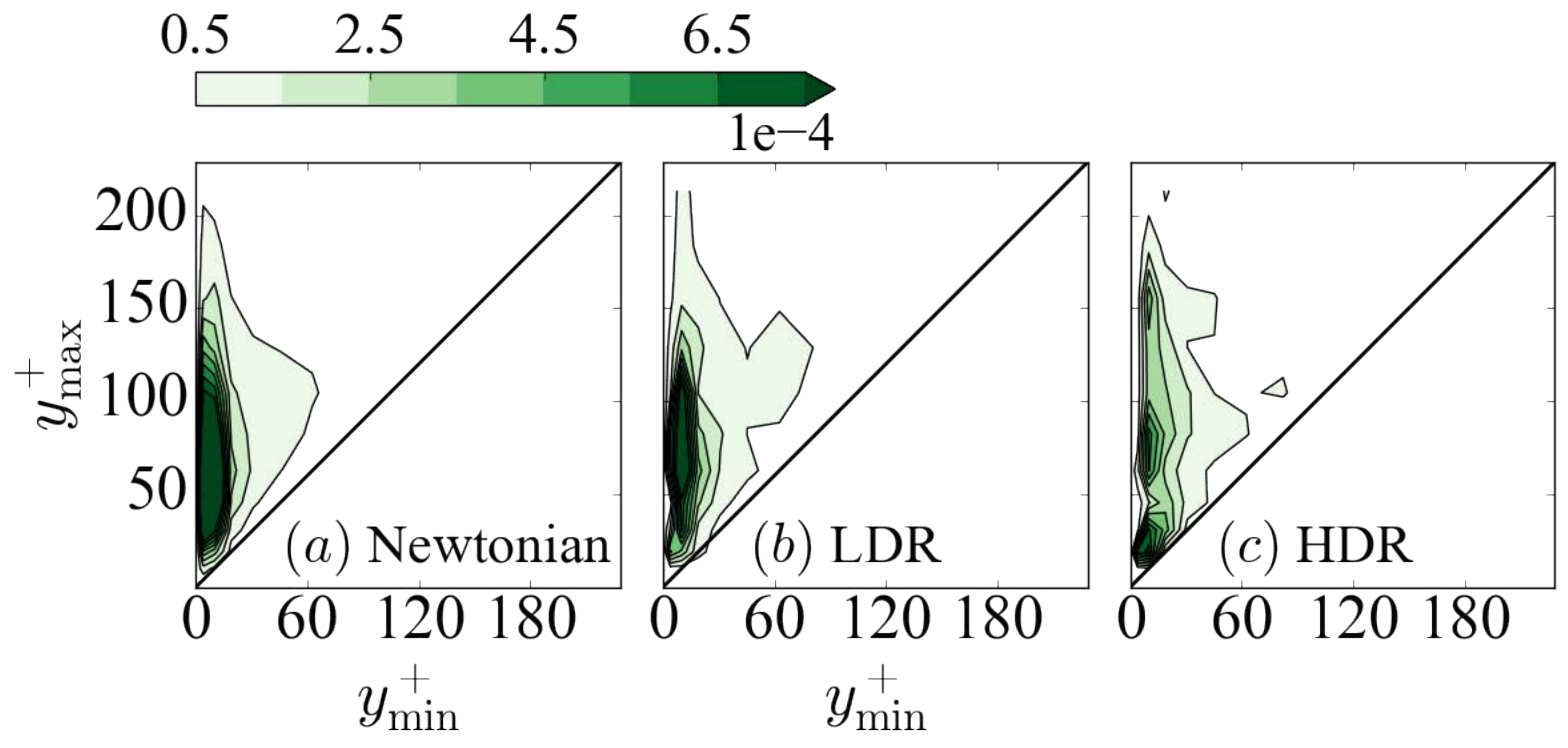}
	\caption{Joint PDFs of the wall-normal positions of the head and tail/legs of hairpin-like vortices (hairpins and branches), as measured respectively by the maximum and minimum $y^+$ coordinates of each vortex axis-line, at different $\mathrm{Wi}$ ($\mathrm{Re}_\tau=400$).}
	\label{fig:vor_height_PDF_cur_re80000}
\end{figure}

Entering HDR does not significantly shift their distribution pattern, despite the substantial reduction in their total count.
Since hairpin-like vortices are products of vortex lift up (generated from lifted quasi-streamwise vortices), suppression of lift up directly reduces the source for their formation.
For those that do come into existence, they maintain their lifted silhouette even at HDR.
The distribution density does decline at HDR, which means the distribution must spread to a wider area owing to the conservation of probability. This reflects an enlarged and more homogeneous boundary layer.
The same joint PDFs for the higher $\mathrm{Re}_\tau=400$ case are shown in \cref{fig:vor_height_PDF_str_re80000,fig:vor_height_PDF_cur_re80000}.
The distribution patterns are again (recall \cref{fig:vor_height_PDF,fig:vor_height_PDF_re80000}) strikingly consistent between different $\mathrm{Re}$. Vortices of the same category are again found in the same wall layer, in inner units, at the two $\mathrm{Re}$ tested.
Reduction in vortex lift up at HDR is consistently observed at both $\mathrm{Re}$.
For quasi-streamwise vortices, the suppression of their lift-up tendency was also observed in the inclination angles of conditionally sampled eddies~\citep{sibilla2005near}.
However, for hairpin-like vortices, which are more predominant among lifted vortices, direct evidence was not previously possible before their axis-lines can be statistically extracted by VATIP.
Since hairpins are most likely generated from the lift-up of streamwise vortices, as conjectured by \citet{robinson1991coherent} and directly observed in DNS by \citet{Zhu_Xi_JNNFM2018}, it is the suppression of the lift-up process itself, not that of any particular vortex type, that is important for interrupting the turbulent momentum transfers between the buffer and upper wall layers and the start of the second stage of DR with distinct log-law region flow statistics.

\section{Summary and conclusive messages}\label{Sec_conclude}
This study focuses on the transition between two distinct stages of DR: LDR and HDR.
Distinction between these two regimes has been made in the literature for two decades because of their different mean flow profiles~\citep{warholic1999influence}. However, it was not until recently that evidences have been established to identify them as two qualitatively different stages marked by a sharp transition in flow statistics and vortex configuration~\citep{Xi_Graham_JFM2010, Zhu_Xi_JNNFM2018}.
For a given $\mathrm{Re}$ and with the introduction of drag-reducing polymers, there are two critical levels of $\mathrm{Wi}$ where two separate mechanisms of DR set in.
The first is the onset of DR: it marks the start of the LDR stage where DR effects are concentrated in the buffer layer.
The second is the LDR-HDR transition where DR effects spread across the log-law layer.

This study leverages the recent development of a new vortex tracking algorithm, VATIP, which enables the automatic detection and extraction of vortex axis-lines without subjective inference~\citep{Zhu_Xi_JFM2019}.
It allows quantitative and statistical analysis of the size, position, conformation, and shape of vortices in a turbulent flow field.
The method is applied to flow fields of a wide range of $\mathrm{Wi}$ covering from the Newtonian limit to HDR.
Vortices extracted by VATIP are then classified using two sets of criteria.
The first is based on the vortex position and lift-up status, which identifies three major groups: (1) type I or attached-flat vortices are closely associated with the wall with little observable lift-up; (2) type II or attached-lifted vortices are generated from the wall but lift up to higher altitudes -- often well into the log-law layer; and (3) type III or detached-flat vortices are similar as type I except that they are found at higher positions with less interaction with the wall (type IV, as discussed above, is not as important and omitted here for the simplicity of discussion).
The second is based on vortex shape which categorize vortices into fragments, quasi-streamwise vortices, hooks (asymmetric or incomplete hairpins), and hairpin-like vortices (the latter further includes canonical hairpins and irregular branches).

Analysis of our DNS results shows that type I (attached-flat) and type III (detached-flat) vortices are nearly all quasi-streamwise vortices, whereas type II contains some quasi-streamwise vortices plus the majority of the curved -- hooks and hairpin-like -- vortices.
Polymers are found to mainly impact attached vortices. At LDR, this effect is an across-the-board weakening of vortex strength without shifting their distribution pattern.
At HDR, polymers start to suppress the lift-up process of vortices and greatly reduces the number of curved vortices including hooks, hairpins, and branches.

A clear conceptual picture thus arises from these observations.
In Newtonian flow, the buffer layer is dominated by flat-lying streamwise vortices. These vortices are prone to lift-up and as the downstream vortex head rises into the log-law layers, it is subject to the impact of transverse flow which can swing and stretch the vortex into a curved contour.
Existence of these highly lifted vortices facilitates the turbulent momentum transport across the wall layers, which is reflected in the well-known log-law flow statistics \citep{townsend1980structure, perry1995wall, lozano2012three}.
At LDR, polymers weaken vortex motion and suppress turbulent fluctuations~\citep{DeAngelis_Piva_CompFl2002, dubief2005new, kim2007effects, kim2008dynamics}, without shifting the overall distribution and balance between different classes of vortices.
As the flow enters HDR, polymers start to suppress the lift-up of streamwise vortices and interrupt the generation pathway of curved vortices (hooks, hairpins, and branches). Reduction in these highly lifted vortices reduces trans-wall-layer turbulent momentum transfer, which offers a clear direction for explaining the changing flow statistics in the log-law layer at HDR.
As vortices become stabilized in the streamwise direction, they become elongated and more detached from the wall. The latter makes them less susceptible to polymer effects.

This is, to our knowledge, the first complete depiction of the vortex dynamics in both stages of LDR and HDR that is based on direct numerical evidences. It substantiates our earlier hypothesis~\citep{Zhu_Xi_JNNFM2018}
\RevisedText{%
which ties the LDR-HDR transition to a fundamental shift of the vortex regeneration mechanism. In that scenario, both Newtonian and LDR turbulence are sustained, to a great extent, by the lift up of vortices which can later burst into fragments and trigger instabilities at streaks elsewhere.
At HDR, since lift up is suppressed and bursting is minimized~\citep{Zhu_Xi_JNNFM2019}, vortices are kept at the streamwise direction and new vortices are more often generated in the immediate vicinity of existing ones.
VATIP analysis results reported in this study are fully consistent with this hypothesized scenario.
Of course, the analysis is still static -- it extracts the conformations of vortices in the flow field without information on their temporal connection. Therefore, it does not directly show the formation and evolution of vortices but rather shows that their conformation statistics match the prediction from the hypothesis.
A dynamical vortex analysis approach, which is as yet not available, will be needed for the direct investigation of vortex regeneration mechanisms.
}%

\begin{acknowledgments}
The authors acknowledge the financial support from the Natural Sciences and Engineering Research Council of Canada (NSERC; No.~RGPIN-2014-04903) and the allocation of computing resources awarded by Compute/Calcul Canada.
The computation is made possible by the facilities of the Shared Hierarchical Academic Research Computing Network (SHARCNET: \texttt{www.sharcnet.ca}).
LX acknowledges the National Science Foundation Grant No.~NSF~PHY11-25915, which partially supported his stay at the Kavli Institute for Theoretical Physics (KITP) at UC Santa Barbara. 
LZ acknowledges the European Research Council H2020 Program (ERC-2014-ADG ``COTURB'') which supported his participation in the Third Madrid Summer School on Turbulence.
\end{acknowledgments}

\appendix

\RevisedText{%
\section{Numerical treatment at the wall boundaries for the FENE-P equation}\label{App_A}
As discussed in \cref{Sec:method:DNS}, when numerically solving the FENE-P equation with AD (\cref{equ_Fene1}), additional boundary conditions are required at the walls.
We follow the procedure of \citet{Sureshkumar_Beris_JNNFM1995} and integrate the FENE-P equation in time first without AD at the walls.
Starting from \cref{equ_Fene1} less the $(1/(\mathrm{Sc}\mathrm{Re}))\nabla^2\mbf\alpha$ term, after taking Fourier transform in the $x$- and $z$-directions and discretization in time with BDAB3, the equation can be rearranged into
\begin{equation}
\frac{\varsigma}{\Delta t}\tilde{\boldsymbol{\alpha}}^{n+1}  = \sum^{2}_{j=0}\left(-\frac{a_j}{\Delta t}\tilde{\boldsymbol{\alpha}}^{n-j}+b_j\tilde{\boldsymbol{N}}^{n-j}_p\right)+\tilde{\boldsymbol{C}}_p.
\label{equ_Fene_discreteNoAD}
\end{equation}
where $n$ denotes the index of the current time step, ``{\large$\,\tilde{\cdot}\,$}'' indicates variables in the Fourier-physical-Fourier space (i.e., no transform yet in the $y$-direction), and $\Delta t$ is the time step size. The numerical coefficients $\varsigma$, $a_j$, and $b_j$ of the BDAB3 method are found in \citet{Peyret_2002}. $\boldsymbol{N}_p$ and $\boldsymbol{C}_p$ group the nonlinear and constant terms in \cref{equ_Fene1}, respectively:
\begin{equation}
\boldsymbol{N}_p\equiv -\boldsymbol{v}\cdot\boldsymbol{\nabla}\boldsymbol{\alpha }+\boldsymbol{\alpha }\cdot\boldsymbol{\nabla} \boldsymbol{v}+\left(\boldsymbol{\alpha }\cdot\boldsymbol{\nabla} \boldsymbol{v}\right)^\mathrm{T}-\frac{2}{\mathrm{Wi}}\frac{\boldsymbol{\alpha}}{1-\mathrm{tr}(\boldsymbol{\alpha})/b},
\label{equ_Fene_Np}
\end{equation}
\begin{equation}
\boldsymbol{C}_p\equiv \frac{2}{\mathrm{Wi}}\frac{b\boldsymbol{\delta}}{b+2}.
\label{equ_Fene_Cp}
\end{equation}
Note that \cref{equ_Fene_discreteNoAD} is explicit as the solution at the future step $\tilde{\mbf\alpha}^{n+1}$ can be calculated directly with information at the current and previous steps ($n$, $n-1$, and $n-2$ steps) only.  
In the case of FENE-P with AD (the full \cref{equ_Fene1}), the time integration equation using BDAB3 is
\begin{equation}
\frac{\varsigma}{\Delta t}\tilde{\boldsymbol{\alpha}}^{n+1} - \tilde{L}_p \tilde{\boldsymbol{\alpha}}^{n+1}  = \sum^{2}_{j=0}(-\frac{a_j}{\Delta t}\tilde{\boldsymbol{\alpha}}^{n-j}+b_j\tilde{\boldsymbol{N}}^{n-j}_p)+\tilde{\boldsymbol{C}}_p,
\label{equ_Fene_discrete}
\end{equation}
where $\tilde{L}_p \tilde{\boldsymbol{\alpha}}^{n+1}$ is the discretized AD term,
\begin{equation}
\tilde{L}_p\equiv \frac{1}{\mathrm{ScRe}}(\frac{\partial^2}{\partial y^2}-4\pi^2(\frac{k^2_x}{L^2_x}+\frac{k^2_z}{L^2_z}))
\label{equ_Fene_AD}
\end{equation}
is the linear operator in the Fourier-physical-Fourier space, and $k_x$ and $k_z$ are wavenumbers in $x$ and $z$ directions.
For each $(k_x,k_z)$ pair, \cref{equ_Fene_discrete} is a second-order differential equation in $y$ solved with the Chebyshev-tau method~\citep{Canuto_Hussaini_1988}. Boundary conditions are required at both walls ($y=\pm1$), for which we use Dirichlet boundary conditions with wall values obtained from solving \cref{equ_Fene_discreteNoAD} at $y\pm1$.
Full details of the entire numerical method for DNS are found in the appendix of \citet{Xi_PhD2009}.
}%


%
%

%


\bibliography{Zhu_bibtex,FluidDyn,Polymer}
\end{document}